# Cybersecurity Threats in FinTech: A Systematic Review

**Danial Javaheri[1*], Mahdi Fahmideh[2], Hassan Chizari[3], Pooia Lalbakhsh[4], Junbeom Hur[1*]**

javaheri@korea.ac.kr, mahdi.fahmideh@usq.edu.au, hchizari@glos.ac.uk, pooia.lalbakhsh@monash.edu, jbhur@korea.ac.kr

1- Department of Computer Science and Engineering, Korea University, Seoul 02841, Republic of Korea
2- School of Business, University of Southern Queensland, Springfield, QLD 4300, Australia
3- School of Business, Computing and Social Sciences, University of Gloucestershire, The Park Campus, Cheltenham GL50 2RH, Gloucester, UK
4- Department of Data Science and Artificial Intelligence, Faculty of Information Technology, Monash University, Clayton, VIC 3168, Australia

* Corresponding author: Junbeom Hur (jbhur@korea.ac.kr), Danial Javaheri (javaheri@korea.ac.kr)

**Abstract**
The rapid evolution of the *Smart-everything* movement and Artificial Intelligence (AI) advancements have given rise to sophisticated cyber threats that traditional methods cannot counteract. Cyber threats are extremely critical in financial technology (FinTech) as a data-centric sector expected to provide 24/7 services. This paper introduces a novel and refined taxonomy of security threats in FinTech and conducts a comprehensive systematic review of defensive strategies. Through PRISMA methodology applied to 74 selected studies and topic modeling, we identified 11 central cyber threats, with 43 papers detailing them, and pinpointed 9 corresponding defense strategies, as covered in 31 papers. This in-depth analysis offers invaluable insights for stakeholders ranging from banks and enterprises to global governmental bodies, highlighting both the current challenges in FinTech and effective countermeasures, as well as directions for future research.

## 1. Introduction

FinTech, an important player in the global economy and financial arena, emphasizes technological innovations aimed at automating financial, a sector notably susceptible to cyber-attacks and cyber espionage. The FinTech ecosystem is characterized by a dynamic, multifaceted network of agents, interacting seamlessly to provide a diverse range of financial products and services to end users [1]. The transformative influence of FinTech on financial services, coupled with the rise of new start-ups and business models, has been so significant that it's termed the 'Fintech Revolution' [2], [3]. Digital threats targeting FinTech services are deemed as potent weapons of mass destruction. The United States classifies significant breaches in this sector as acts of war, meriting retaliatory actions, as highlighted in an Allianz[1] report. However, technological evolution presents a paradox. On one hand, it equips legitimate sectors with sophisticated, but on the other, it arms malevolent actors with enhanced tactics, leading to advanced cyber-attacks [4], cyber espionage [5], the creation of deep fakes [6], and potential misuse of sophisticated AI constructs such as ChatGPT [7]. The rapid rise of FinTech can be attributed to the mutual integration of cutting-edge technologies, in particular AI, neural networks, cloud/edge computing, Blockchain (BC), and embedded systems. This fusion, however, has also positioned FinTech as an important playground for different IT-based and AI-enabled technologies and applications, some leveraged as avenues for malicious activities and cyber threats [8], [9]. Based on X-Force Threat Intelligence Index 2023[2] report, the *Finance and insurance* sector ranked second among the industries most targeted by cyber criminals since 2018.

FinTech utilization in small and medium enterprises (SMEs) during the COVID-19 pandemic is the most crucial facet in the growth of FinTech and the strength of SMEs. Major banks worldwide have been seriously trying to develop and implement their own FinTech platforms or extend their collaboration with other relevant FinTech start-ups to change their banking landscape from traditional systems to the modern era [10]. Although FinTech-powered services have provided ease of access and high availability, they have caused new cybersecurity concerns, in particular for banks and businesses. For example, from the beginning of Feb. 2020 to the end of Apr. 2020, attacks on the financial sector surged by 238% globally, with 80% of financial institutions reporting an increase in cyber-attacks, according to Modern Bank Heists 3.0 report from VMware[3].

Considering this alarming number of attacks against FinTech, the necessity for systematic and comprehensive analysis on the topic increasingly arises to deal with the potent attacks and mitigate their increasing threats [11]. To pave this way, our survey investigates the most potential cyber threats against FinTech and looks into the most effective strategies against them. Although some previous works have discussed the cyber threats against FinTech, they lack studying new cyber threats that emerged in the past three years and offering defensive solutions. Also, none of them has considered threats originating from human mistakes and paid less attention to procedure-based threats.

---
[1] https://www.agcs.allianz.com/news-and-insights/expert-risk-articles/cyber-attacks-on-critical-infrastructure.html
[2] https://www.ibm.com/downloads/cas/DB4GL8YM
[3] https://www.vmware.com/learn/security/modern-bank-heists-2020.html

To our best knowledge, this study is the first work that has investigated new cyber threats that emerged after 2020; considered threats from all possible sources, i.e. technology, humans, and procedures; and offers defensive strategies to deal with these threats as well as put them into comprehensive comparison. Aiming to bridge gaps in previous surveys, this survey responds to the following research questions (RQs).

RQ1- What are the most recent cyber threats against FinTech systems?
RQ2- What cyber threats are more sophisticated and destructive?
RQ3- What are existing defensive strategies to tackle these threats and what threats can be mitigated by a certain defensive strategy?
RQ4- How to adopt and/or implement defensive strategies in FinTech systems to respond to potential threats in the real world?
RQ5- What lessons can be learned from previously occurring cyber-attacks?
RQ6- What are the future research directions to deal with cyber threats in FinTech?

Our contribution can be summarized as (a) investigating the most significant security threats in FinTech, ranking them based on impact and severity by reviewing 43 of 74 papers, (b) finding the most effective solutions to neutralize/mitigate these threats by examining 31 of 74 papers, (c) presenting a novel taxonomy of security threats within FinTech, and establishing a hierarchy for defensive measures, (d) conducting a comprehensive comparison of the current security threats, assessing their impacts, and evaluating defensive strategies based on their effectiveness and technical details, (e) highlighting existing security vulnerabilities in today's FinTech systems and suggesting future research directions to address these shortcomings.

## 2. Related Work

Studies in the independent realm of FinTech and cybersecurity are abundant; however, only a few papers have reviewed cybersecurity threats and their implications against FinTech.

Gai *et al.* [12] conducted a survey on FinTech challenges, in particular, Security and Privacy (S&P), by reviewing contemporary achievements. They proposed a theoretical data-driven framework, named DF2, looking into a broad range of data security techniques in FinTech from three dimensions, i.e. Business operations, Outsourcing, and Fin-privacy. In this survey, four technical perspectives were considered for discussing S&P challenges, which are (*i*) hardware and infrastructure, (*ii*) data techniques, (*iii*) service models, and (*iv*) applications and management. This work focused on analyzing the practical applicability of defensive technologies, e.g. *fully homomorphic encryption* (FHE), to neutralize insider/outsider attacks in FinTech, leading to the conclusion that an up-to-date and precise awareness of FinTech is urgently needed for both industry and academia.

Another systematic review was performed on publicly available reports by Huang *et al.* [13] to indicate how cyber-attacks and cybercrime have become organized against businesses and the financial sector from a value chain standpoint. This includes existing, evolving, and emerging threats. The authors have recognized 24 critical value-added activities and the correlation among them by focusing on the commercialization, specialization, and cooperation for cyber-attacks. They also extracted a framework and *cybercriminal value chain model* for understanding the financial sector's cybercriminal ecosystem. Considering six factors, *Ease of attack, Potential benefit, Ease of benefit realization, Psychological costs, Expected penalty cost, and Operational costs*, in different case studies, they assessed the effectiveness of the proposed framework in developing secure financial systems able to tolerate novel adversaries attack. Eventually, this study concluded with recommendations for encouraging collaboration between different sectors to react against the attacks and targeting cybercrime control points like hacker training/recruiting communities.

Mehrban *et al.* [14] reviewed existing open challenges with FinTech. They showed that current FinTech is associated with many sensitive issues, such as security threats, cyber-attacks, and privacy concerns in the financial sector. They reported that security incidents are practiced in institutions and state organizations that offer IT-based financial services more than any other sector. Among their findings, cyber-attacks were recognized as the most crucial challenge in the future of FinTech. Aiming to make FinTech more secure, the authors analyzed the most recent cyber-attacks, privacy concerns, and existing detection tools. They further proposed a taxonomy of current technology-based cyber threats and security solutions for FinTech, such as risk detection, risk reduction, authentication and access control, data storage and processing, and data usage cycle. They also gave some anticipation about future security threats in the financial industry.

Table 1 compares previous surveys and our systematic review from various points of view. There are also studies looking into cybersecurity challenges in FinTech focusing on specific technologies such as Blockchain [15]. Moreover, for general surveys on FinTech and cybersecurity, we refer the readers to [16] and [17], respectively.

**Table 1:** A comparison between related surveys and our systematic review.

| # | Comparison aspect | Ref. [12] | Ref. [13] | Ref. [14] | This paper |
|---|---|---|---|---|---|
| 1 | Year of survey | 2017 | 2018 | 2020 | 2023 |
| 2 | Technology-originated threats | ✓ | ✓ | ✓ | ✓ |
| 3 | Human-originated threats | ✗ | ✗ | ✗ | ✓ |
| 4 | Procedure-originated threats | ✓ | ✗ | ✗ | ✓ |
| 5 | Taxonomy or hierarchy | ✓ | ✓ | ✓ | ✓ |
| 6 | Attack detection methods | ✗ | ✗ | ✓ | ✓ |
| 7 | Defensive solutions | ✓ | ✗ | ✗ | ✓ |
| 8 | Future research directions | ✓ | ✓ | ✓ | ✓ |

As indicated in Table 1, none of the previous related surveys has considered all technology, human, and process-originated security threats. Besides, the latest surveys return to 2020; hence, the new threats that emerged after 2020 have not been investigated. To our best knowledge, this paper is the first that studies security threats from all sources, including technology, humans, and procedures, and meanwhile analyzes, ranks, and discusses the most effective strategies to deal with them. This paper also offers recommendations from real-world experience in mitigating security threats in FinTech.

## 3. The Survey Methodology and Statistics

This survey was conducted based on the PRISMA methodology [18]. The following query was used to retrieve papers from ten well-known scientific libraries i.e. ScienceDirect, IEEE Xplore, Springer Link, ACM Digital Library, Taylor and Francis, MDPI, Wiley, IET Digital Library, Emerald, and Hindawi.

*"(cyber\*) AND (threat\* OR attack\* OR crime\*) AND (financ\*) AND (Business\* OR Industr\* OR Organization\*OR Institution\* OR "Enterprise" OR Bank\*)"*

To conduct this survey, we have followed the steps of PRISMA methodology, i.e. 1- Identification, 2- Screening, 3- Eligibility, and 4- Inclusion. After running the aforementioned query on the ten designated libraries and filtering the output to a time window from 2015 to date, 3620 papers were identified (Step 1). We considered an 8-year timeframe, aiming to focus on the most recent studies as cyber-attack patterns are changing and evolving rapidly and to maintain the length of this survey appropriate for readers. Removing the duplicate studies by checking the titles, 2109 papers remained in total (Step 2). After that, by manual inspection through reading the abstract, we found 512 papers eligible for full-text reading (Step 3). Ultimately, by full reviewing the papers remaining from Step 3, 74 papers included for qualitative synthesis (Step 4) of which 68 studies are journal articles and 6 records are from conferences.

The inclusion criteria by manual inspection during Step 3 and Step 4 were (*i*) published by leading venues, a Q1/Q2 WoS-indexed journal or an A\*/A/B CORE Ranking conference, aiming to stay focused on more important studies, (*ii*) availability of validation in the paper - empirical papers rather than white papers, (*iii*) written in English, (*iv*) exactly related to the RQs mentioned in Section 1. The relevancy was validated by reading the abstract in Step 3 and full reviewing of the papers in Step 4 to ensure the candidate paper has either addressed a cybersecurity threat in the FinTech sector or proposed a solution to deal with an already discovered threat. To distinguish these 74 papers included for qualitative synthesis from general references that were added to support claims or further delineation of threats/defense models, their list with publishing venues was reported in Table A-1, Appendix.

Fig. 1 illustrates the diagram of the PRISMA methodology, including all four Identification, Screening, Eligibility, and Inclusion steps, along with the number of included/excluded papers (n) at each step. Furthermore, to provide a better resolution, the number of papers in Step 3 (papers eligible for full-text reading) and Step 4 (papers included for qualitative synthesis) for each year from 2015 to 2023 has been indicated in Fig. 2.

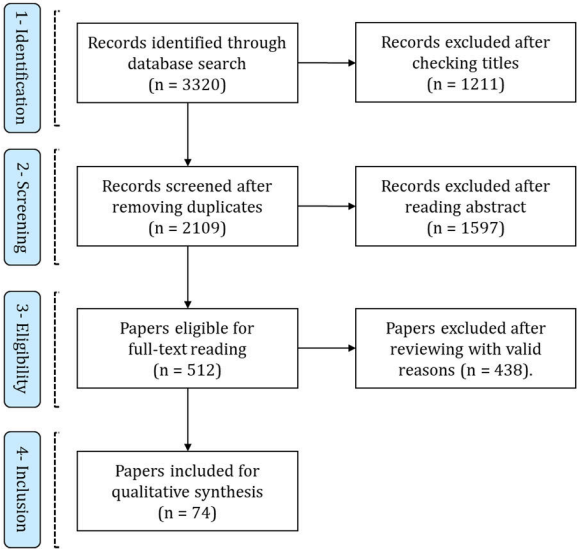
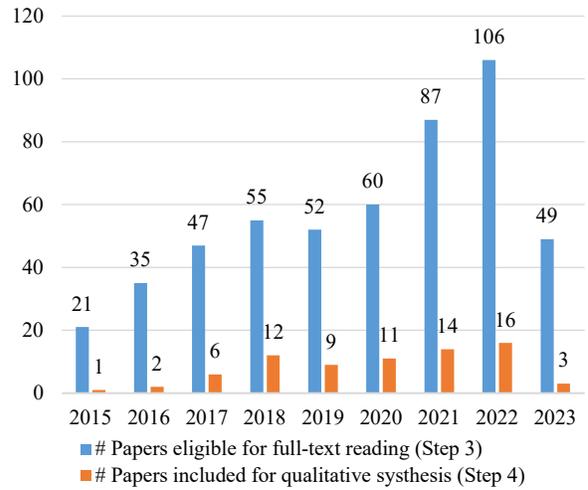

**Fig. 1:** The steps of the PRISMA methodology in our survey where (n) indicates the number of included/excluded papers at each step.

**Fig. 2:** The number of papers eligible for full-text reading (Step 3) compared to the number of papers included for qualitative synthesis (Step 4).

As shown in Fig. 2, there has been an ascending trend in the number of published papers related to FinTech security since 2015, indicating the topic has become more critical and attracted the attention of more researchers. Fig. 3 indicates the number and percentage of papers included for qualitative synthesis (Step 4) for each library. As indicated in Fig. 3, with 60.8% (45 records), the majority of papers on this topic were published by ScienceDirect, followed by IEEE Xplore with 14.9% (11 records).

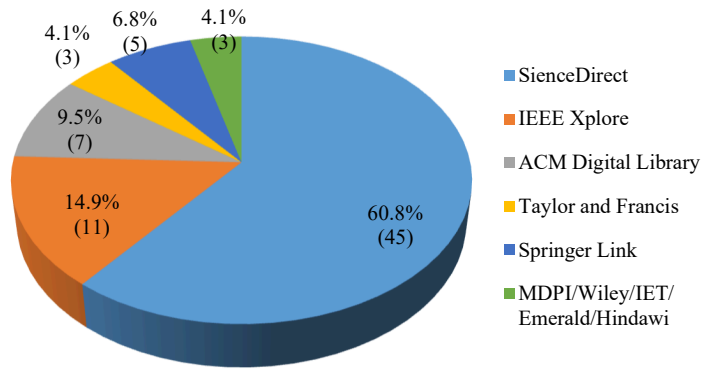

**Fig. 3:** The number and percentage of paper included for qualitative synthesis (Step 4) for each library.

To provide a novel taxonomy of new cybersecurity threats against FinTech and corresponding defensive methods, we first manually classified the papers included for qualitative synthesis (Step 4) into two 'Threats' and 'Defenses' which are studied in Section 4 and Section 5, respectively.

## 4. Implications of Potential Security Threats

After perusing each of the papers included for qualitative synthesis (Step 4), we manually labeled and classified them into 3 classes: technology-based, human-originated, and procedure-related, aiming to evolve the previous classifications. A resilient cybersecurity framework as a control against cybersecurity threats in the financial services and socio-technical system was proposed by Ambore *et al.* [19] in which the human element (either as the originator, the medium, or the actual executor) was applied to the former technology-based taxonomy of cyber-threat landscape presented by European Union Agency for Network and Information Security (ENISA)[4]. Furthermore, William

---

[4] https://www.enisa.europa.eu/publications/etl2015

Stallings in his textbook [20], has considered humans as an essential attack surface, in addition to technology (hardware/software). Meanwhile, in the definition of Vulnerability in the *Glossary of Key Information Security Terms* published by the US National Institute of Standards and Technology (NIST)[5], weaknesses in security procedures and internal controls have been underscored as an attack surface. Although some studies have emphasized the role of policies and procedures on specific requirements for cybersecurity, this attack surface has not been well-addressed so far. To bridge this gap and to our best knowledge, our survey is the first that has included all three technology-based, human-originated, and procedure-related elements to present a novel and refined taxonomy of threats and corresponding defenses in the FinTech sector.

## 4.1 Technology-based Threats

Papers have been further subdivided into 11 subclasses after running a Latent Dirichlet Allocation (LDA) topic modeling and classification on the whole text of 43 papers already labeled as 'Threats' manually. The model has classified the papers based on the relevancy and closeness of their topics, the distance between words, and the frequency of similar words. Ultimately, we tuned the titles for each subclass from a cluster of words and expressions offered by the model and assigned them to each of the three main technology-based, human-originated, and procedure-related groups, according to our knowledge of the paper.

The majority of cybersecurity threats in FinTech originate from technology flaws, misuse, and misconfiguration, or using technology to deceive individual victims, which are explained in the following 6 subsections.

### *4.1.1 Malware Attacks*

*Financial Services* has absorbed 19% of the total number of malware attacks and has seen about 18.3 billion dollars of loss merely in 2017, putting it the top sector targeted by malware, followed by *Utilities and Energy* sector, according to the reports published by Evolve[6].

Although spyware families, including keyloggers and screen recorders, have posed an extreme threat to FinTech-enabled systems, some classes of malware have been dedicated designed to target financial services, called financial-targeted or banking malware [21], [22]. Carbanak, GCMAN, Dridex, Shylock, and Zbot have been recognized as the most destructive instances of financial-targeted or banking malware. They are responsible for targeting many banks and financial institutions for illegitimate money transferring over recent years [23]. After illegitimate money transferring, hijacking credit card credentials from POS devices using memory scraping was the second dominant attack seen in banking malware, followed by robbing crucial corporate information to get ransom from the victims [24]. Fig. 4 indicates the dominant percentage and era that such malware programs have been active considering financial services, according to reports published by Kaspersky[7]. As shown in Fig. 4, Zbot has been the most dominant banking malware over recent years; however, Ramnit has taken the lead from Zbot since 2022.

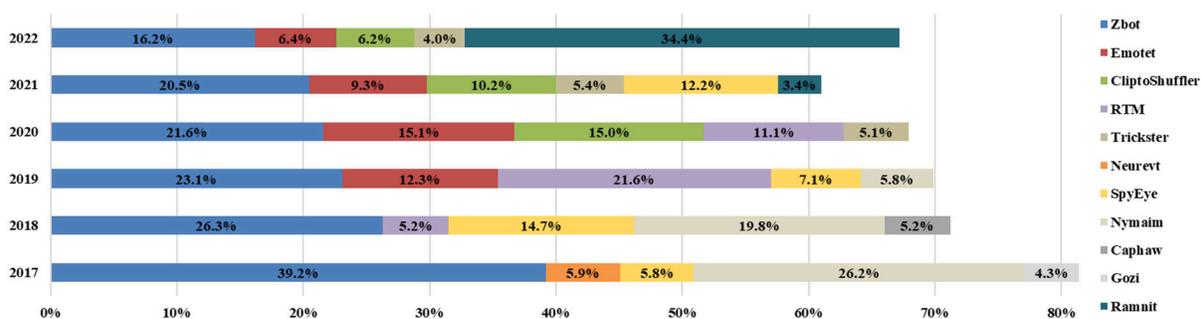

**Fig. 4:** Top-five dominant banking malware in recent years.

Malware production has experienced significant annual growth in recent years; in contrast, the number of banking malware has remarkably decreased, as shown in Fig. 5, according to data published by AV-ATLAS[8]. Upon examining the underlying causes, we discovered that the prevalence of banking malware has not actually decreased; rather, these malicious programs have evolved to become more covert, thereby eluding detection by Anti-virus scanners (AVs). The root cause includes sophisticated obfuscation techniques and metamorphic engines used by novel banking malware to evade Anti-virus detection, as well as the adoption of Advanced Persistent Threat (APT) tactics that involve executing attacks in a deliberately slow and low-key manner. APT classes of malware split the attack procedure into several steps with time gaps between them. Therefore, AVs cannot correctly identify the correlation between features associated with malicious behavior. This is because indicators are distributed among various malware pieces/processes [25]. Novel banking malware tries to swipe small sums from a vast array of bank accounts (a feature of being low) in a way that owners and/or auditors, even forensic analysts, often fail to detect the loss. By repeating this procedure over an extended period across a diverse customer base (a feature of being slow) malware developers have accumulated significant financial gains.

---

[5] NIST IR 7298 Revision 2, https://nvlpubs.nist.gov/nistpubs/ir/2013/NIST.IR.7298r2.pdf
[6] https://evolvemga.com/financial-loss/
[7] https://securelist.com/financial-cyberthreats-in-2022
[8] https://portal.av-atlas.org/malware/statistics

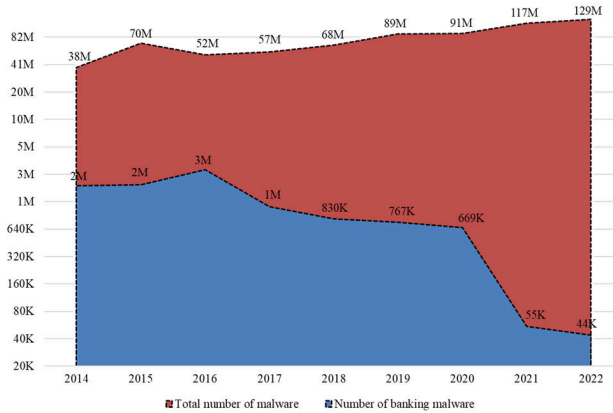

**Fig. 5:** The number of banking malware against the number of total malware produced in recent years.

Peris-Lopez and Martin [26] have disclosed a significant vulnerability in virtual keyboards employed by a large number of banks across the world to deal with the threat of keyloggers. By implementing a proof-of-concept hardware-based trojan (HT) able to record the user's passwords and PIN codes, they proved such virtual keyboards in e-banking applications are highly vulnerable to shoulder surfing and malicious software-based attacks, particularly trojans. Their adversary model was embedded on the board of a VGA/LCD display controller core with such a negligible overhead of resources, i.e. 5.6% of registers and 4.3% of Slices, that cannot be detected by visual inspection, as shown in Fig. 6, we barely could find just one small area - indicated in Yellow - that differs from a clean VGA core. This HT is able to reveal private information each time the user clicks on the virtual keyboard using a GologicTM logic analyzer to recognize the corresponding digits. This HT was evaluated as extremely harmful as it can defeat the defensive countermeasures, such as digit obfuscation or keyboard mutation, used by e-banking applications to deal with common keylogger programs.

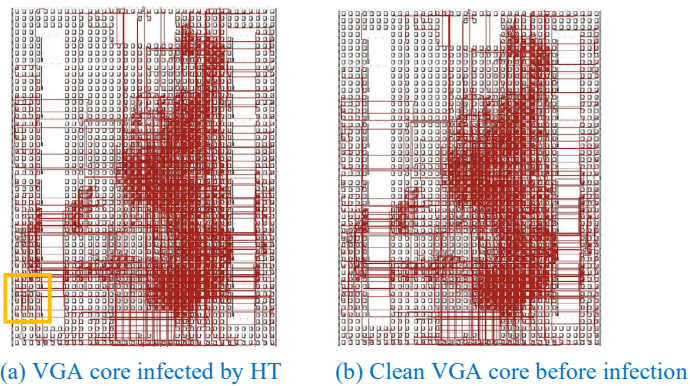

(a) VGA core infected by HT  (b) Clean VGA core before infection

**Fig. 6:** The layout of an infected VGA core (a) on a Spartan-3E Xilinx FPGA compared to the clean one (b), where 681 registers and 752 Slices were used in the infected core against 645 registers and 721 Slices in the original clean core and the number of LUTs was same in both cores, adapted from [26].

*4.1.2 DDoS Attacks*

Although numerous comprehensive studies have identified denial of service (DoS) and distributed denial of service (DDoS) attacks, timely detection of such attacks continues to pose a significant challenge to commercial targets and high-profile government bodies. HSBC bank and Irish National Lottery fell victim to the largest-ever DDoS attacks. They witnessed coming in at 500Gbps of illegitimate traffic, which has been away from malicious traffic of other types of cyber-attacks by far [27]. For instance, Low-Rate DDoS (LR-DDoS) attacks, a specific subcategory of DDoS, are notoriously difficult for Intrusion Detection Systems (IDS) to identify and for firewalls to neutralize [28]. This type can be categorized as one of the APTs, causing issues in communication chains and infrastructures and can lead to service outages and network collapses [29]. Such failures are among the most damaging events in terms of financial loss and reputation for targeted sectors. The study referenced in [30], based on extensive interviews with key stakeholders in the Swedish financial sector, suggests that DoS attacks represent the most likely threat facing the financial industry.

Recent efforts have aimed to tackle the persistent issue of LR-DoS attacks. In a recent research study by Pérez-Díaz *et al.* [28], a new IDS system based on machine learning techniques has been proposed to detect LR-DDoS attacks. In this IDS, six machine learning models, mainly tree-based classifiers, including RepTree, Random Tree, Random Forest (RF), J48, Multi-Layer Perceptron (MLP), and Support Vector Machine (SVM), were put into experiments to detect LR-DDoS attacks. The model was built using data from the Canadian Institute of Cybersecurity and, when tested on an Open Network Operating System (ONOS) controller within a Mininet virtual environment, demonstrated a peak detection accuracy of up to 95%. Despite such efforts, the risk of LR-DDoS attacks remains a critical concern for the financial sector since state-of-the-art methods are still unable to detect LR-DDoS attacks accurately and reliably. The existence of merely a 5% possibility of occurring LR-DDoS attacks can potentially disrupt essential financial services, translating to considerable financial and reputational losses for the industry.

Bandwidth Distributed DDoS (BW-DDoS) [31] is another form of DoS attack wherein the attacker overwhelms the target with excessive traffic volume, thereby impairing its ability to handle legitimate requests. BW-DDoS attacks also differ in terms of the protocol used to flood the target, in which a privileged zombie with full control over its host, is a highly effective attack agent who can deliver fake IP packets.

Flooding attacks are the most common subcategory of DDoS attacks against cloud host FinTech services. Among flooding attacks, the session initiation protocol (SIP) is a text-based application protocol used to establish, control, and terminate multimedia sessions on voice-over IP (VoIP) signaling protocols. Like other Internet protocols, deployment in live scenarios showed its vulnerability to flooding attacks. These attack patterns are similar to those against TCP protocol; however, it has appeared at the application level of the Internet. An attacker can trigger an *INVITE* flooding attack by sending a vast stream of *INVITE* messages consisting of different session IDs like *Call-Id, From, or To* aim to drain the SIP memory since the latter has to store the connection state information to ensure reliability for a designated time period. For instance, in the case of an invalid URI in the SIP message, the proxy should forward it to another domain and keep its state information or a copy of the message for 30 seconds or longer. Other SIP messages, such as *Ack*, *Bye*, *OK*, might also be utilized to carry out flooding attacks by sending numerous messages that do not belong to any session. Considering the vast employment of VoIP authentication mechanisms by many banks, this attack aims to make the bank server so busy that it can no longer respond to authentication requests coming through the VoIP channel, besides caller ID spoofing attacks to bypass the whole authentication mechanism in mobile/telephone banking systems [32], [33].

Reflection-based attacks [34] represent another prevalent form of DDoS attack, in which uncompromised systems are co-opted to direct a massive influx of traffic toward the targeted entity, thereby exhausting its network bandwidth. The advantage of reflection-based attack strategies is that they allow attackers to covertly channel traffic to the targeted system while helping the originating attack host remain undetected. In this scheme, the attacker sends IP packets that carry the victim's IP address in the source address field. When the server receives the request, it responds to the victim rather than to the actual source, making detection more difficult. Smurf attacks are a specific type of network layer DDoS attack in which a large volume of Internet Control Message Protocol (ICMP) packets, each with a spoofed source IP pointing to the intended victim, are sent across a computer network using an IP broadcast address. In amplification DDoS attacks [35], the attackers control a group of slave and master zombies and coordinate them to flood a huge number of requests into reflector systems. Botnets may be employed by attackers to intensify these reflective attacks, further obscuring the attacker's identity. Amplification of Reflective DDoS attacks [36] is a subtype of DDoS that augments the victim's reflected traffic through certain permitted protocols. These attacks are particularly complicated because they generate a higher number of response messages than the original requests from the attacker, leading to an exponential increase in incoming data that overwhelms the victim's resources. Protocols that amplify traffic, such as DNS or Network Time Protocol, are commonly exploited to execute these amplification attacks [35]. For a clear understanding, we have classified different DDOS attack variants and highlighted those particularly challenging for FinTech, as shown in Fig. 7.

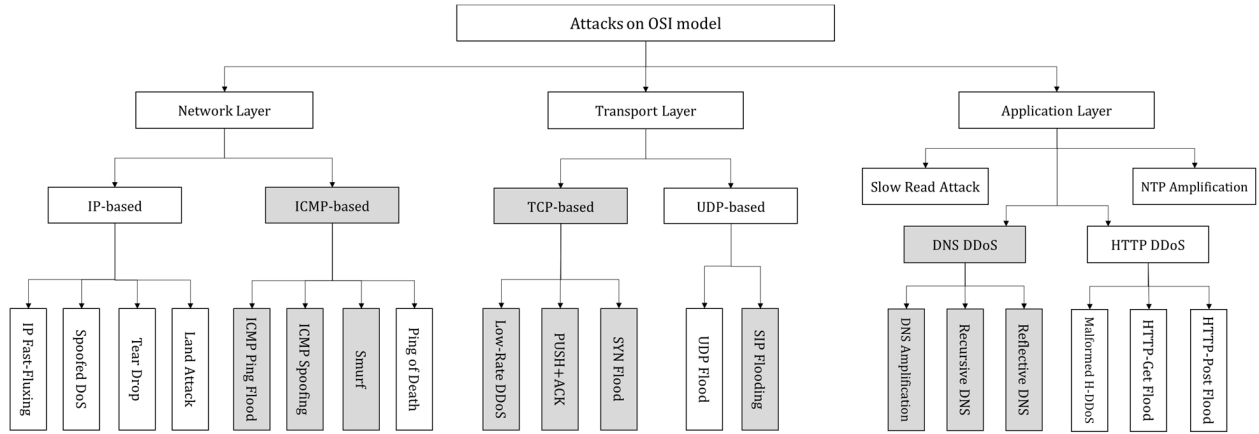

**Fig. 7:** A hierarchy of DoS variants where those that have been more challenging for FinTech were highlighted.

### 4.1.3 Digital Extortion

Digital extortion, particularly through ransomware, has recently surged as a major cyber threat to businesses, from SMEs to large enterprises that host FinTech services. These attacks not only cause immediate financial losses but also adversely affect accessibility, productivity, and reputation, making them an escalating concern for both governmental and business entities. This type of cyber-attack has occurred more against the financial sector than other sectors due to the higher data sensitivity level in financial systems [37]. Notably, the WannaCry and Petya ransomware attacks resulted in estimated damages of $8 billion and $10 billion, respectively [38]. Although many victims employ robust anti-ransomware and Anti-malware applications, these attacks were not entirely preventable. This underscores the current limitations in defense strategies against ransomware threats, promoting intensified efforts to address these vulnerabilities in recent years.

Sherman *et al.* [39] explored the limitations of current supervised learning methods in detecting ransomware variants. These limitations mainly stemmed from difficulties in extracting the inherent traits and concealed origins of various attack patterns in these new ransomware variants, often a result of obfuscation, code encryption, and polymorphism.

Alternative defensive strategies concentrated on data backup, given that emerging ransomware can evade detection using stealth techniques such as executing a series of pre-attack API calls named 'Paranoia' which mislead AVs and determine a potential time for execution [40]. Specifically, Min *et al.* [41] devised a device-level backup strategy that doesn't require additional storage space. Their method uses a hardware accelerator that runs content-based detection algorithms to identify ransomware attacks at high speed paired with a precise backup control mechanism that reduces data backup space overhead. Using a Microsoft solid-state drive (SSD) simulator and a prototype on the OpenSSD platform, their evaluation demonstrated the solution's efficacy in countering ransomware with high accuracy and minimal performance overhead. Furthermore, their approach also reportedly defends against ransomware classes that target and erase backup data.

From the industry perspective, some leading cybersecurity firms such as Kaspersky, Bitdefender, and McAfee have developed specialized anti-ransomware applications to detect and counteract ransomware threats, as addressing this type of malware often demands a dedicated ML classifier. Despite these advancements, these attacks persist, resulting in significant financial and reputational damage daily.

### 4.1.4 Industrial Espionage

In today's competitive landscape, major producers and service providers across various industries face a surge in cyber-industrial espionage incidents. This competition is exacerbated by the swift integration of information technology in contemporary businesses, especially within the financial domain. The main motivation of notorious players is to reap financial benefits from cyber-enabled industrial espionage (IE), thereby undermining competitors' market positions. IE is an umbrella term that covers a broad spectrum of malicious activities aimed at securing competitive edges. Unfortunately, nowadays, new technologies such as Data Science and the Internet of Things (IoT) have notably amplified the intensity and frequency of IE activities [42].

Sadok *et al.* [43] identified insider attacks as among the most significant cyber threats and explored technical countermeasures. Their research indicated that a combination of social and technological approaches can provide a more effective solution to responding to insider attacks due to the nature of this type of cyber-attack. They also highlighted the importance of fostering a security-aware culture, bridging the gap between solution designing and implementation across various entities.

Härting *et al.* [44] indicated that susceptibility to espionage isn't solely based on business size but rather on its technological and innovative capacities. This study reveals that SMEs, given their typically limited resources – human, financial, and technical - are particularly vulnerable to industrial espionage, often resulting in undesirable knowledge leakage. Also, this study emphasized the fact that it's crucial to differentiate economic espionage from other industrial espionage forms, as the initiators and the espionage's nature vary based on whether it's state-sponsored or stems from the private sector. The influences of different constructs in conjunction with the IE threat on SMEs can be extracted from this study. Thirteen variables were designated and categorized into four aggregated determinants, i.e. motives, significance of trade secrets, prevention, and security risks, as indicated in Fig. 8. We further analyzed these variables and highlighted those that are applicable to the financial sector and the security of FinTech systems.

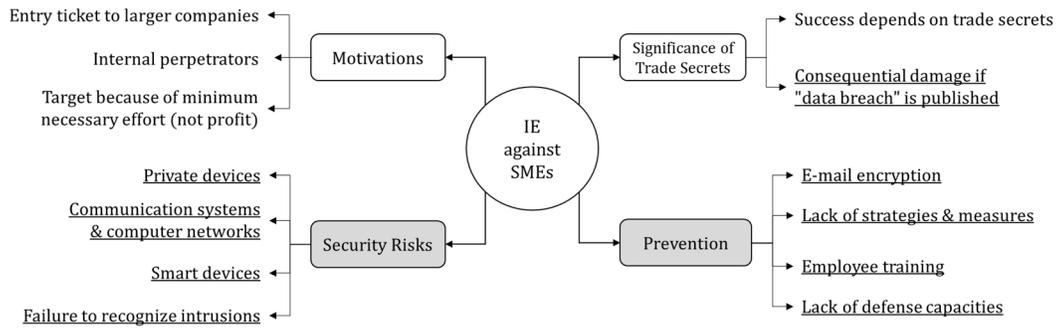

**Fig. 8:** The constructs of IE threats on SMEs where variables applicable to FinTech are underlined, adopted from [44].

It is worth noting that IE is associated significantly with insider attacks, making this attack more severe and complicated. Malicious insiders can inflict significant damage, providing adversaries access to resources, establishing vulnerabilities, or leaking sensitive information. Therefore, an in-depth understanding of IE is desperately vital in today's increasingly competitive global business landscape.

### *4.1.5 Vulnerabilities in Industrial Internet of Things (IIoT)*

The gigantic number of IoT devices is continuously and rapidly increasing, leading to the interconnectedness and interoperability of systems in numerous critical systems [45]. Modern banking systems widely employ this technology and devices to carry out financial affairs more accurately and rapidly than traditional methods. In the same way that IoT technologies enable cyber and physical interaction between systems, they also enable sophisticated attack vectors against vital systems. Hence, security has demonstrated vital importance in IoT due to the emergence of many new threats, such as DDoS attacks, escape attacks in the virtualization layer, as well as vulnerability exploitation attacks in the software layer. In this context, there are many common vulnerabilities between IIoT and standard IoT devices, which have caused them to be assailable to attackers [46].

Advanced Persistent Escaper (APE), as a state-of-the-art threat, performs a cross-VM escape attack in the virtualization layer, in addition to a cross-layer penetration in IIoT devices. This highly sophisticated attack can barely be detected even by cutting-edge technologies, reported by Sha *et al.* [47]. They also suggested a method to detect cross-VM and guest-to-host escapers at the virtualization layer. However, due to the possibility that a portion of attacks remained undetected, they ultimately offered to use an Identity-based Broadcast Encryption system as a self-defense mechanism to secure the communications between different modules in the virtualized environment.

Besides APEs, some novel classes of malware generated by advanced obfuscator engines [48] are specifically designed to penetrate IIoT devices, leading to a rapidly growing threat in this area. Furthermore, botnet infection has propagated to IoT devices and plays a significant role in a wide range of malicious activities, particularly exploiting these devices to perform DDoS attacks. The most widespread cyber-attack that exploited the distributed nature of IoT devices was carried out by Mirai malware. Mirai is a botnet active on IoT devices, which is responsible for several catastrophic DDoS attacks against many with a significant portion of the Internet lockdown on October 21, 2016, when millions of users lost their access to over 1200 important websites, including Netflix, Twitter, and several international Banks, for about one entire day [49]. Using forensic analyses of Mirai's source code provided by [49], [50], we depicted the attack pattern of Mirai malware shown in Fig. 9.

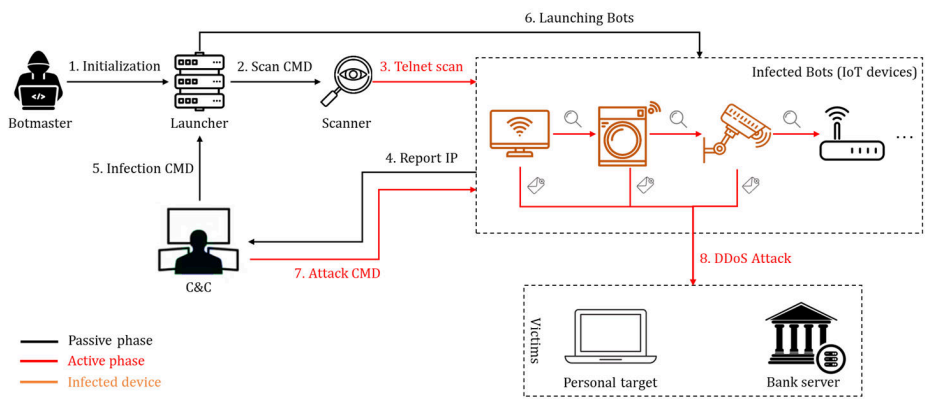

**Fig. 9:** The attack pattern of Mirai Botnet.

As illustrated in Fig. 9, Mirai malware initializes a DDoS attack by triggering its *Launcher* through Botmaster at the first step. Then, the Launcher sends a scanning command to the scanner application, responsible for commencing a Telnet scan on IoT devices targeted as bots (Step 2). During this scan in Step 3, vulnerable devices are determined as candidates to be compromised. Next, each penetrated device sends a message to the malware's C&C and reports its own IP (Step 4). After that, the C&C decides to either compromise the candidate Bot or seek a more prone candidate. This step iterates until a sufficient number of devices get compromised and a network of bots (Botnet) is created. Having a robust Botnet, C&C sends an infection command to the Launcher (Step 5); afterward, the launcher missions the bots to start the attack against a certain victim - targeted by the C&C - by generating and sending network packets, Step 7-8. Since the packets come from heterogeneous IoT devices distributed worldwide, neither the victim's IDS systems can properly detect nature and trace the source of the attacks, nor can its firewalls block the malicious incoming traffic. Receiving the packets from different routes by the victim causes an overflow in its incoming gateway bandwidth or computational/storage resources, leading to system failures. Such system failures resulted in service outages and have imposed severe financial loss and reputation damage - in some cases not compensable - to the financial sector and many household name banks.

Jiang *et al.* [51] have measured the surfaces of attack concerning IIoT devices in a network. In this research, several concrete attack scenarios, including some variations of the Mirai malware tailored for attacking industrial equipment [52], were implemented to gain the system's control actuators and impose several hazardous actions. In another study, Madan *et al.* [53] indicated that Android is not the only OS used in IoT devices; several distributions of Linux have widely been employed as the OS for IIoT devices. Hence, the malware designed for the Linux OS needs to be considered another source of threat. Perusing the existing detection and classification methods, analysis techniques, and available tools for dealing with Linux-based IoT malware, this research concluded that plenty of efforts and research need to be conducted to bridge the existing security gaps between the theory of IIoT and its real-world applications.

Taheri *et al.* [54] reported the lack of sufficient IIoT malware samples as the most severe challenge to training a DL-based model to detect and classify such malware programs. To overcome this barrier, they proposed a framework consisting of a server-side and a client-side module. On the client-side, a generative adversarial network (GAN) has been employed to generate the required data for performing poisoning attacks dynamically. At the same time, the server-side module monitors the whole system and trains a collaboration model by feeding anomalies aggregated via the GAN network. This method guarantees that devices can efficiently communicate without privacy concerns.

Stellios *et al.* [55] found that the threat of IoT-enabled attacks is highly associated with cyber-physical systems. Using risk-based analysis methodology, this study illustrated attack tree topologies where the tree's root indicates a critical system built recursively according to the recognized cyber-physical system interactions. Well-known building blocks, such as Common Vulnerability Scoring System (CVSS) and Common Vulnerabilities and Exposures (CVE), were used for threat modeling, indicating multi-hop attack patterns, and assessing the risks of IoT-enabled attacks against critical infrastructures like financial technology.

### 4.1.6 Power Attacks

The emergence of smart grids and virtual power plants, in which power consumers become producers while offering numerous advantages and facilities, have paved the way for cyber-attacks against the energy sector. In the past few years, these attacks disrupted a wide range of services in the financial industry, resulting in millions of dollars of financial loss, as well as terrible reputation damage. According to our survey, this type of attack, as an alternative, occurs when the attackers cannot directly target a financial system. The aim is to interrupt financial services and transactions by losing power supplies.

The consequences of this type of attack are so massive that they are highly employed in modern cyber warfare, e.g. cyber-attacks against the Ukraine power grid in December 2015 resulted in millions of dollars in financial loss [56]. It makes maintaining a power system operational a vital task. Han *et al.* [57] evaluated cyber-attacks against an electric power organization in South Korea for a four-year window. Their study included the examination of IPs and raw data, which led to disclosing that about 95% of these attacks originate from outside the country. As a result, considering six different features, i.e. real login, blocking complexity, foreign relation, external relation, stopping impact, and stopping tolerance, they suggested a priority to restrict external IPs not associated with a foreign business for vital infrastructure.

*Measurements* and *critical parameters* are the most valuable targets for attackers against smart grids. The *critical parameters* are a set of model parameters, the errors of which cannot be recognized due to the lack of local *measurements*. Attacks that target *measurements* should be carried out in real-time. In contrast, attacks against network parameters can be conducted offline as the attacker merely needs to alter the parameters once. Such alternations are typically executed via code and false data injection attacks, propelling the systems into a state of error [58]. False data injection attacks have the potential to disrupt the power system state estimation process, leading to state estimators providing incorrect information to system operators. This can culminate in economic or physical harm to the power infrastructure [59]. The consequences of inaccurate estimations extend beyond the power system themselves, potentially resulting in catastrophic outcomes for electricity consumers. This includes the suspension of both Alternative Current power flow model and Direct Current supplies essential to critical clients such as banks, leading to substantial financial losses. Thus, ensuring the security and stability of power plants becomes an intricate endeavor. Despite the implementation of various security measures, cyber-attacks have still managed to cause large-scale blackouts and financial damages within smart grid systems [60]. Fig. 10 presents a schematic representation of a virtual power plant (VPP), highlighting the strategy an attacker might use to disrupt financial services by cutting the power connection between banks and VPP infrastructure. Such link disruptions can be effectively achieved through false data injection attacks targeting VPPs [61].

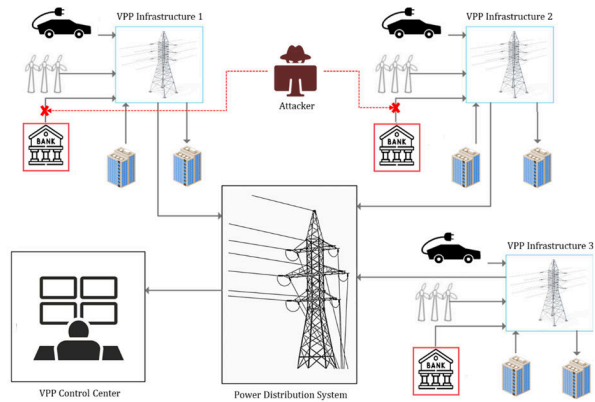

**Fig. 10:** VPP power supply system where the attacker tries to disrupt bank services via false data injection attacks, adopted from [61].

Furthermore, the injected errors by the attacker will always remain undetectable because the criticality property of parameters is entirely independent of the system's operating points, making detecting and confronting cyber-attacks against critical parameters more complicated, studied by Xu *et al.* [62]. They also modeled the attacks against critical parameters based on a bi-level programming problem, seeking the best attack pattern or the worst case the system operator can encounter via simulations for injecting undetectable errors into critical parameters. Security researchers have tried to mitigate the effects of power attacks against the financial sector, but it still is a severe challenge with many open problems.

## 4.2 Human-originated Threats

The second subclass of threats stems from human error or poor practices. Unfortunately, human misbehavior, particularly lack of knowledge, has significantly exposed FinTech systems and their clients to a broad spectrum of vulnerabilities. Compared to the advanced cyber defense tools available today, humans represent the most fallible element within a digital ecosystem, where social, cultural, and psychological factors can remarkably affect their decisions. Hence, targeting humans is often the most accessible and least challenging gateway for those seeking unauthorized access to secured environments.

### *4.2.1 Insider Attacks*

Insider threats originate from malicious employees and have always been a persistent concern for businesses, companies, and organizations. Over recent years, the number and scale of attacks carried out by insiders have rapidly increased in various domains, particularly in financial services. Previously, cybersecurity solutions in the financial sector concentrated on defending against external attackers and intruders, while the number of attacks established from inside was more than the number of outsiders. Fig. 11 depicts and compares the rate of insider attacks against outsiders during the last decade, according to *M-Trend* report released by FireEye[9] in 2023.

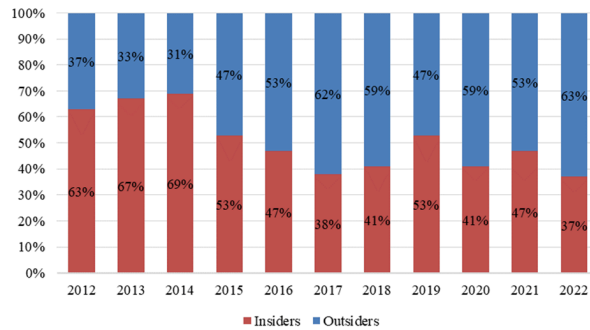

**Fig. 11:** The percentage of attack sources in the last decade.

As shown in Fig. 11, insider attack poses a significant portion of cyber-attacks. Besides, this report revealed that ransomware attacks are the most common cyber threat from insider sources. This gap has attracted the interest of many cybersecurity researchers to bridge it, so some debates have broadened for threat management and paying more attention to insider attacks, including trusted employees, contractors, and business partners [63]. Although plenty of efforts have been made by researchers to bridge this gap, the effectiveness of such solutions and tools has scarcely been indicated in real environments due to the carelessness of the owners, leading to numerous cyber-attacks against businesses and the financial sector from insiders. In this regard, Erola *et al.* [64] have cooperated with three international financial-related companies to assess a proposed anomaly detection system. Their system was fed with real data for over a year to determine performance constraints for insider threats and intrinsic weaknesses in the operational contexts and validate the results. Then, the authors reported the outcomes of deploying the detection system in real infrastructure, including the limitations, experiences, and lessons learned.

Another challenge in dealing with insider threats is that ML-based detection methods, which heavily rely on feature engineering, can barely allow a precise distinction between the behavior of malicious insiders and benign users. This is because of numerous obstacles related to the characteristics of data in this area, including the lack of labeled insider threats, sparsity, heterogeneity, complexity, and the adaptive nature of insider threats. Further, novel deep learning (DL) approaches, particularly artificial neural networks (ANNs), offered a new paradigm to train end-to-end models from complex data, providing higher accuracy and performance than traditional ML models to detect insider threats. Nonetheless, applying deep learning models to advance the insider threats task still faces many challenges and restrictions, such as dealing with adaptive attacks and the lack of labeled data, as deep learning models are data-hungry to be well-trained [65]. Meanwhile, manual analysis by security experts is impossible for entities with many employees. Therefore, insider attacks are categorized as one of the most potent and dangerous threats against the financial sector, and much more effort must be made to detect and confront these threats, as human behavior can be much more sophisticated and obfuscated than machines.

---
[9] https://www.mandiant.com/resources/blog/m-trends-2023

### *4.2.2 Social Engineering Attacks*

Social engineering attack is a multi-faceted concept considering physical, technical, social, and socio-technical dimensions, which cannot be smoothly detected and blocked using technical protection measures. There has been a substantial movement in the way social networks and cloud apps are developed, from monolithic programs that contain all functions in a single codebase to massive graphs of single-aspect loosely-coupled microservices [66]. The improvements in the secure implementation of digital infrastructures and innovative cyber-defense strategies created a shift from targeting the system to targeting humans using these systems [67]. As the weakest links in information systems (IS), operators, technicians, and system users play a crucial role in fulfilling social engineering attacks. In these attacks, the victim is not a machine, library, or code. The victim is a human, and the attacker uses social strategies to gain trust and run the malicious plan. People are accessible through their emails, social media accounts, etc., and all these access points can be harnessed to deploy and implement a social attack vector [68]. The increasing adoption of *bring your own device* (BYOD) policies within both private sectors and business environments, including banks, coupled with communication over third-party channels have broadened the spectrum of attack vectors susceptible to sophisticated social engineering tactics. When combined with the exploitation of zero-day vulnerabilities, social engineering escalates into a formidable tool frequently deployed by advanced persistent threat actors targeting major financial institutions [69].

The 2019 pandemic event, which is still around, is very relevant to social engineering attacks, where working in home offices using personally available devices and communication channels became a new normal. This attack aims to embarrass, threaten, and harass targeted entities, which can negatively affect the victims' reputations. In these new geographically dispersed workplaces, the security of the communication infrastructure, which is generally cloud services, is taken for granted, and an account profile or a trusted email address could be enough to create trust. This mishap cannot be fixed by strengthening the security of the infrastructure of the communication channel [70]. Unfortunately, social engineering methods for compromising information system security are not just a theory or a potential threat. There have been many successful attacks against companies like Facebook [71], PayPal [72], Toyota [73], Shark Tank [74], Sony Pictures [75], and Uber [76].

Krombholz *et al.* [69] presented a taxonomy of social engineering attacks and classified them based on operators, channels, and types of attacks in different scenarios, as well as advanced attack vectors used in modern communication technologies. These scenarios were gathered by examining real-world samples of successful social engineering attacks against giant techs, big enterprises, and banks. We further improved and refined this taxonomy by including Doxing attack vector, which was a missing piece. Doxing refers to collecting information about entities, including people and companies, by use of social media, search engines, and other publicly available information sources, which is the initial step for further steps and harsher attacks, such as espionage and fraud [77]. With this explanation, the new refined taxonomy of social engineering attacks is indicated in Table 2, where the attack scenarios and vectors that have taken place more frequently in the FinTech industry were asterisked.

All attack scenarios are articulated based on human psychology and even customized for a specific character, mainly the person who has the piece of knowledge the attacker is trying to obtain. The attack can be initialized easily as a simple email, a phone call, or Doxing techniques, which have been reported as the most prevalent media to perform a social engineering attack [78]. It can be elaborated through socio-technical approaches to surf on human curiosity and infiltrate information systems using infected road apples [79]. Spear-phishing attacks are another sophisticated type of social engineering attack that targets a small group of individuals and are built around their personal and environmental profile. These attacks are tailored according to relevant datasets and results from data-mining strategies to increase the success rate of the attack.

**Table 2:** A refined taxonomy of social engineering attack scenarios and vectors - an improved version of [69].

| Attack scenarios | | Attack vectors | | | | | | | |
|---|---|---|---|---|---|---|---|---|---|
| | | Doxing | Phishing* | Waterholes | Baiting* | Shoulder surfing | Dumpster diving | Reverse social engineering* | APT* |
| Operator | Human* | ✓ | ✓ | | ✓ | ✓ | ✓ | ✓ | |
| | Technology (software/hardware) | ✓ | ✓ | ✓ | | | ✓ | ✓ | ✓ |
| Type | Social* | ✓ | | | | | | ✓ | |
| | Technical | | | ✓ | | | | | ✓ |
| | Socio-Technical* | ✓ | ✓ | ✓ | ✓ | | | ✓ | ✓ |
| | Physical | | | | | ✓ | ✓ | | |
| Channel | Social Media | ✓ | ✓ | | | | | ✓ | |
| | Instant Messenger | | ✓ | | | | | ✓ | |
| | VoIP* | | ✓ | | | | | ✓ | |
| | Email* | | ✓ | | | | | ✓ | ✓ |
| | Website* | ✓ | ✓ | ✓ | | | | | ✓ |
| | Physical | ✓ | ✓ | | ✓ | ✓ | ✓ | ✓ | |

\* Indicates those attack scenarios/vectors that have occurred more frequently against the FinTech sector.

When evaluating the social engineering taxonomy concerning the financial industry, approaches might be interpersonal or non-interpersonal as the communication channels between attackers and victims. The interpersonal technique via non-electronic means typically involves direct face-to-face contact between the attacker and a particular victim. The threats of impersonation, shoulder surfing, and reverse engineering are included in this section. Via human abilities such as friendliness, compliance, and sympathy, an attacker can covertly get financial credentials using a variety of tactics. Most social engineering attackers try to avoid face-to-face contact with their victims; instead, they prefer to use electronic means such as email, the Internet, and other digital media to abuse human nature, betray the victim, and accomplish their goals. Baiting, website spoofing, pretexting, and phishing are some of the most common digital social engineering attacks [80], [81].

### *4.2.3 Internet Fraud*

Although the rapid emergence of modern banking has allowed numerous facilities, including remote trading, it has also attracted fraudsters to utilize the advancement of technology for their malicious goals. Internet Fraud has resulted in a loss of billions of dollars to service providers and customers in the world's financial sector every year, highlighting the severe need for fraud detection systems. Button *et al.* [82] investigated the statistics of Internet Fraud (IF) against theft according to the Crime Survey of England and Wales (CSEW) for the past decade. According to their findings, a 50% drop in police-recorded theft and the concurrent 63% drop in CSEW theft indicate that it is becoming less popular as an illicit way of acquisition. Conversely, recorded fraud has more than doubled, and CSEW fraud has surpassed theft, implying that acquisitive criminality has evolved, shifting from theft to fraud.

Current fraud detection methods detect IF as an abnormality from the regular purchasing routines of customers [83]. In research by Ali *et al.* [84], the anatomy of fraud in various user-centered technologies was illustrated. This research indicated that the anatomy of Internet fraud against banking systems consists of three perspectives, i.e. technology-enabled frauds, mechanisms, and tools for applying fraud attacks, and detection/prevention systems on traditional telecommunication, mobile, and the Internet. Based on this anatomy, they also suggested some countermeasures to prevent fraud attacks.

Despite the efforts that have been made to improve the accuracy of fraud detection systems, there is a significant gap in the accurate detection of banking fraud, as reported by Carminati *et al.* [85]. This study has studied the role of decision support systems in detecting banking fraud. This research found that fraud detection in the banking sector cannot be well-accurate without user-centric approaches to assessing each customer's spending habits. This study also showed simulation tools are desperately needed for running proofs-of-concept codes to measure the accuracy of banking fraud detector models and prove their performance.

During this survey, we realized that, unfortunately, the number of published works that addressed this threat was not considerable, so there is an obvious need for investigating this cyber threat against the financial sector, especially for those attackers who are aware of the internal structure and mechanism of the system.

## 4.3 Procedure-related Threats

This subsection studies the third and last subclass of security threats, those that arise from inadequate or improperly implemented internal policies within banks and financial institutions. It also considers the threats due to flawed procedures among entities operating in the financial sector, and eventually, the mishandling of data maintenance.

### *4.3.1 Data Leak*

Data leaks can be caused by simple human or technical errors, but the impact on governments, companies, and users can be catastrophic. Like social engineering attacks, data leaks might result from users' mistakes. In this situation, they cannot be considered a consequence of poor cyber-security strategies or a lack of equipment. Organizations must create customized security policies and pieces of training according to their organizational architecture and cultural characteristics. These policies might even differ from department to department. Along with the evolution of the organization, data leak security policies must be updated and evolved through effective security and shared awareness pieces of training [86]. Also, it is believed that the data should be treated as it deserves, where the greater the secrecy, the more likely the data is to leak; it is the lesson that can be learned from the WikiLeaks phenomenon. To prevent data leak, countermeasures should be applied against insider threats, which are the major cause of data leaks. Also, the security of cloud services should not be taken for granted, as we saw several incidents of data leaks with Amazon, Google, etc. [66], [87]. Eventually, as Shannon's information theory reminds us, data is almost ultimately prone to leak, and to address that, the focus should be on the weakest links, which are human beings [86], [88]. One of the biggest incidents in this sector was the massive data breach concerning Equifax. In that, the personal and credit card information of about 148 million clients had been compromised [24], or millions of records leaked as T-Mobile was breached for the sixth time in four years [89].

### *4.3.2 Data Breach*

In a study by Lee *et al.* [90], a routine activity theory (RAT) was introduced to design a risk assessment model based on staff's perceptions of risks relative to potential breaches of sensitive data in several financial institutions. The authors empirically evaluated indicators like potential targets, motivated offenders, and the influences of managerial affairs. The results of this survey indicated that accessibility of targets, inertia, and perceptions of value have the most impact on the evaluation of risks of data breaches in banks and financial institutions. This study also extends RAT to be accountable for determining the amount of sensitive data prone to be breached among online information. Ultimately, they concluded that data breaches and data leaks in banks and financial institutions could cause significant damage to organizations.

## 4.4 Comparison of Threats

This subsection presents a meticulous comparison between investigated threats against FinTech. The comparison consists of various standpoints, including the nature of attacks, motivations, the level of danger/destruction, consequences, and technical details, as shown in Table 3.

In Table 3, the level of destruction was measured based on the amount of financial damage caused by each threat, according to the published reports. The level of danger was determined according to the frequency of the attack's occurrence during the window of this survey, and the other technical information was derived from papers studied in this survey. Then, we sorted the attack based on the combination of destruction level, danger level, motivations, as well as an attack surface. Furthermore, Fig. 12 illustrates a timeline of the most destructive cyber-attacks against FinTech and their damages in USD since 2015, according to data published by CARNEGIE[10].

---

[10] https://carnegieendowment.org/specialprojects/protectingfinancialstability/timeline. Accessed 2023

**Table 3:** A comparison of threats against FinTech perused in this paper.

| Rank & name of threat | References | # Papers | Targets | Motivation | Threat impact | | APT | Main platforms | Manual/Automate |
|---|---|---|---|---|---|---|---|---|---|
| | | | | | Attack Surface | Destruction level | | | |
| 1- Malware Attacks | [21], [24], [26] | 3 | Operating systems | Massive destruction | ◐ | ● | ✓ | Network, Portable devices | Automate |
| 2- DDoS Attacks | [28], [32], [33], [35] | 4 | Network Infrastructure | Service disruption, Reputation damage | ◐ | ◕ | ✗ | Network | Automate |
| 3- Digital Extortion | [37], [39], [40], [41] | 4 | Humans | Financial gain | ◐ | ◐ | ✗ | Web | Hybrid |
| 4- Industrial Espionage | [42], [43], [44] | 3 | Industrial control systems | Technology thief | ◔ | ◐ | ✓ | ICS systems | Automate |
| 5- Vulnerabilities in IIoT | [46], [47], [49], [50], [51], [52], [53], [54], [55] | 9 | Edge devices | Espionage, sabotage | ◔ | ◐ | ✗ | IoT devices | Automate |
| 6- Power Attacks | [57], [58], [59], [60], [61], [62] | 6 | Power suppliers | Financial gain/reputation damage | ◐ | ◕ | ✗ | Power distribution network, Smart grids | Automate |
| 7- Insider Attacks | [64], [65] | 2 | Anything | Revenge, Financial gain, and Espionage | ◐ | ● | ✓ | N/A | Manual |
| 8- Social Engineering | [69], [70], [77], [80], [81] | 5 | Employees | Information gathering | ● | ◕ | ✓ | Social media | Manual |
| 9- Internet Fraud | [83], [84], [85] | 3 | Humans | Financial gain | ◐ | ◐ | ✗ | Web | Hybrid |
| 10- Data Leak | [86], [87], [89] | 3 | Humans | Information gathering | ◐ | ◐ | ✓ | Covert channels | Hybrid |
| 11- Data Breach | [90] | 1 | Humans | Information gathering | ◐ | ◕ | ✗ | Web | Hybrid |

● very high ◕ high ◐ medium ○ low

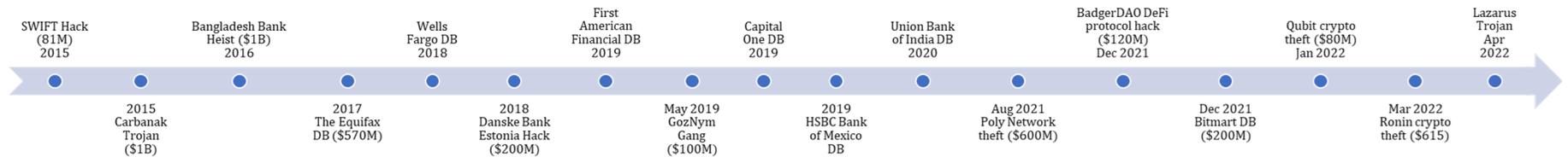

**Fig. 12:** A timeline of the most destructive cyber-attacks against FinTech since 2015.

## 5. Results of Reviewing Defense Strategies

Upon applying LDA topic modeling on 31 papers corresponding to 'Defenses', they were classified into 9 subclasses according to the relevancy and closeness of their topics and words discovered and validated by the model. Following a process similar to the one used for 'Threats', a thorough review was undertaken. Subsequently, the 9 defense strategies were manually labeled and assigned into three technology-based, human-originated, and procedure-related groups.

### 5.1 Technology-based Approaches

This subsection presents the findings from our review of technology-based defense strategies, which encompass a range of methods, tools, and applications. These were synthesized from 23 defense papers that were classified in this subclass. Each identified strategy is capable of addressing one or more of the threats outlined previously in Section 4.

#### 5.1.1 Anti-malware and Intrusion Detection

Behavioral features and indicators need to be extracted and modeled to determine whether a program is malicious. Banking trojans are well-equipped with obfuscation techniques, including metamorphic and packing engines, to distort behavioral features and mislead AV scanners, firewalls, and IDS systems. To extract latent behavioral features from modern obfuscated and intelligent classes of malware, a precise analysis approach with several considerations is desperately needed. Two main categories of malware detection methods include traditional signature-based methods and behavioral analysis methods [91]. Behavioral analysis methods are classified into two static and dynamic approaches. The static approach extracts features from malware's file when stored on a hard disk - as a file - without any execution [92]. In contrast, the dynamic approach executes malware binaries to extract features from its running process [93]. Fig. 13 compares these approaches from four aspects, i.e. scanning speed, depth and accuracy of analysis, vulnerabilities against malware's obfuscation techniques, and burden of implementation.

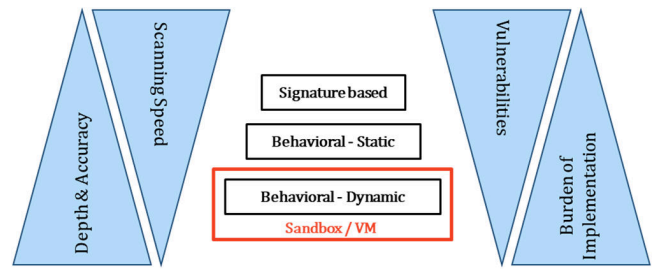

**Fig. 13:** A comparison between malware detection methods.

As indicated in Fig. 13, signature-based methods provide the highest scanning speed but the lowest depth and accuracy. These methods are highly vulnerable to malware obfuscation techniques, which means they can simply be deceived by banking malware. For example, when a malware program inserts junk code among its executable sections to change its structural signature, it becomes much more challenging to detect by signature-based methods [94]. On the other hand, detection methods based on dynamic behavioral analysis provide the highest depth of analysis and accuracy. They also have the most minor vulnerabilities when confronting obfuscated and metamorphic malware. However, implementing such strategies is complex since these methods need an isolated environment, such as a VM or Sandbox, for the risk-free execution of malware. Furthermore, the scanning speed of these methods is low compared to the other methods. Dynamic malware detection methods work based on the API name or frequency of API calls as the features for behavioral modeling and detecting if a malicious process is running on the system. However, these features are insufficient to detect banking malware as they can mislead these detectors using dynamic code loading, runtime code generation, code injection, puzzling, and other obfuscation techniques [95]. Signature-based and static analysis approaches are inappropriate for dealing with modern obfuscated banking trojans. A hybrid strategy, which utilizes the advantages of both static-behavioral and dynamic-behavioral analysis methods simultaneously, offers more chances to detect banking trojans.

Bai *et al.* [96] developed an Anti-malware system named *DBank*, to identify banking trojans designed for Android OS. They extracted APIs from a dataset collected from VirusTotal[11] and applied graph mining techniques to detect Banking trojans. *DBank* adopts a hybrid analysis approach and, according to the evaluation, it can identify banking trojans and distinguish them from good-ware accurately - up to 99.9% area under the curve (AUC) and 0.3% of false positive rate. The performance evaluation was conducted on a data-driven analysis of Android Banking Trojans (ABT), including Asacub, BankBot, FakeToken, Marcher, and Svpeng, and compared to 63 AVs available on VirusTotal. Whereas this system is highly accurate in detecting Banking Trojans, it conducts a binary classification, and its performance was not evaluated against other banking malware classes.

---
[11] https://www.virustotal.com/

Dassouki *et al.* [97] introduced a dynamic and adaptive approach designed to safeguard cloud-hosted financial services from SIP flooding attacks. This is achieved through the analysis of temporal characteristics and fingerprints of message exchanges. The main advantage of this system for the FinTech industry is that it was evaluated on an extensive set of experiments employing geographically distributed virtual machines (VMs) in cloud service providers. The testing framework comprised nine clients from different countries and a server positioned in a separate location. The SIP clients interfaced with the Internet using the Java Jain-SIP library operating on Linux Ubuntu within the Amazon AWS cloud environment. The server was built based on a SIPP[12] traffic generator running on a Linux Ubuntu VM hosted on Digital Ocean, as indicated in Fig. 14 (a). Since AWS allows dedicated connections to route inter AWS regions, locating the main server on Digital Ocean forces the incoming traffic from clients to employ public Internet to reach the main server. In this system, the threshold for attack detection is dynamically tuned considering the acceleration of an imminent attack, as shown in Fig. 14 (b). Once the detection system raises a signal, the prevention system blocks the malicious incoming traffic. It acts based on a threshold of abnormal events and a history made of the session's fingerprints instead of endpoints IPs and URLs, which consume less computational resources and less restriction when filtering attack traffic.

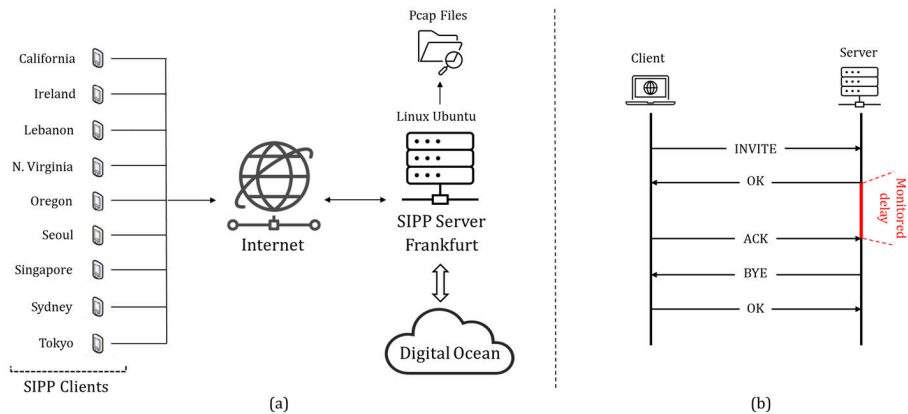

**Fig. 14:** (a) A holistic view of the testbed, (b) exchanged messages in each call of cloud-based attack scenarios, adapted from [97].

The system succeeded in resisting large-scale emulated flooding attacks launched from mutually distant data centers. This geographical scalability in attack detection and prevention, as well as low detection time and low sensibility to false alarms, are well-aligned with the international banks needs for Intrusion Detection and Prevention Systems (IDPS) to mitigate DDoS attacks in general and Telephony Denial of Service (TDoS) in particular, considering the intrinsic nature of global banks to offer 24/7 financial services for numerous clients worldwide.

### 5.1.2 Access Control Systems

Misconfiguration of access control systems might have led to many dynamic threats from outsiders and insiders, imposing risks to privacy protections due to massive data sharing with third parties. This concern is put to an extreme when financial data is shared. Some data protection models have been presented so far to guarantee the privacy of customers' information.

Malaquias and Hwang [98] conducted an empirical study on trust assessment as an essential factor in modern banking systems. They conducted a Structural Equation Modeling and Confirmatory Factor Analysis on a set of 1077 questionnaires collected in Brazil in 2016. Their study demonstrated that a sort of information asymmetry can be addressed to build trust in the FinTech systems, accelerate its adoption, and thus the efficacy of banks. Meanwhile, they reported a negative relationship between trust in FinTech systems and undergraduate courses in technology. To address this barrier and build trust, a comprehensive hierarchy of security and privacy (S&P) solutions on FinTech platforms was proposed by Hernandez *et al.* [99], as shown in Fig. 15.

---
[12] https://sipp.sourceforge.net/

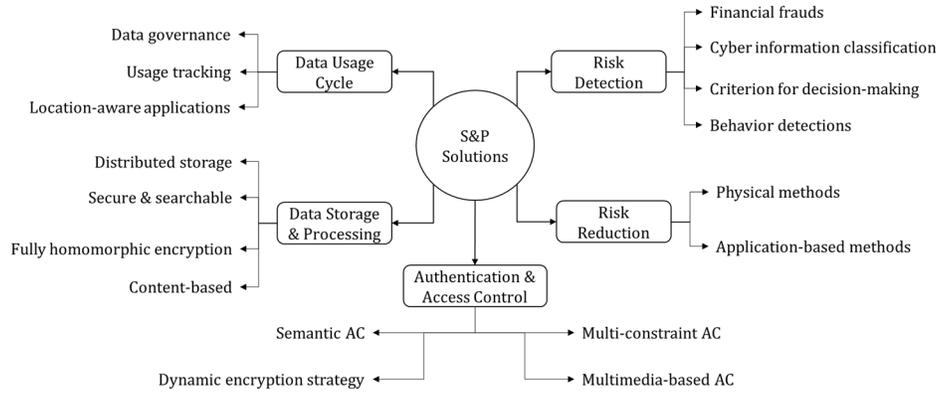

**Fig. 15:** S&P solutions for data protection on FinTech platforms, adopted from [99].

Among S&P solutions, access control (AC) mechanisms have a pivotal role. However, a limited number of these AC schemes are dedicated to financial services and FinTech platforms. As one of the state-of-the-art models, Qiu *et al.* [100] employed attributed-based access control (ABAC) to enable the proactive protection of financial customers' privacy information and a data self-deterministic scheme aiming to guarantee that unanticipated parties cannot reach the privacy data. This scheme employed proactive determinative access (PDA) and Attribute-based Semantic Access Control (A-SAC) algorithms to constrain data access and prevent users' data from unwanted operations using a user-centric approach. The outcomes of this research can guarantee the privacy-preserving financial data of customers in banks and financial institutes up to an acceptable extent.

Today's market-based technologies, especially FinTech, need to specify adjustable limitations on the power of attackers, admit systematic monitoring and the uniform identification of attacks, and allow verifiable responsibility for cyber-attacks. AC systems used in the financial sector require specialized instrumentation to regulate the trade and use of access rights in banks' databases. The importance and sensitivity of financial data necessitate novel and customized approaches to AC that generic systems cannot fully provide. Therefore, domain controlling and fine-tuning the exposure to attacks are required in response to emerging risks, on-time detection of intrusions, minimizing resource unavailability, and statistical analysis of currency flaws [101]. It is evident that access control in FinTech systems is more complicated than the other domains and severely needs a customized model well-tuned for the intrinsic nature of financial data.

### 5.1.3 Blockchain Technology

A growing number of scientific and practical domains are eagerly interested in leveraging Blockchain (BC) as a disruptive technique to empower IS in many applications in FinTech, IoT, and token economy [102]. A survey by Lei *et al.* [103] profiled 443 articles published between 2016-22 to present a new classification of BC applications and highlight the promising role of BC in the systems designs of the future IS. This classification includes three phases for IS innovation, i.e. (*i*) system design, (*ii*) implementation, and (*iii*) impacts, containing 188, 114, and 68 articles, respectively. However, 73 articles could not be categorized into any of these categories. Among their reports, we noticed that the finance sector and financial technology had attracted a considerable number of published papers, 38 articles in total.

Employing BC, cryptocurrencies have appeared in a special place in the free economy. Besides the significance and economic characteristics of cryptocurrencies, the technical aspects of this field are also remarkably considerable [104]. Using BC as a data store infrastructure in some financial-related applications instead of traditional database systems has promoted data privacy, as well as transaction integrity, by separating transactional data from personally identifiable information [105]. According to this reference, BC awareness can potentially reduce privacy and security concerns with FinTech.

Although BC has offered many advantages for banking systems and FinTech, particularly minimizing fraud possibility, several new challenges appeared when facing big data. Nowadays, for any active player in the financial sector, connecting to an immense network such as the Internet and social media is not a choice but a necessity to survive in the market, resulting in hasting the production of Big Data, particularly *business-generated data*, with seven V's features, namely *Volume*, *Velocity*, *Variety*, *Valence*, *Veracity*, *Variability*, and *Value* [106]. This issue negatively affects the adoption of BC technology for FinTech. Besides, new vulnerabilities, including crypto-mining, crypto-jacking attacks, and uncertainty in the cryptocurrency market, have appeared by using the BC as the infrastructure for the financial sector. Also, there is a concern with user anonymity in all cryptocurrencies since they might be compromised for illegal financial flows and criminal activities. Thus, de-mixing algorithms were proposed to mitigate this concern by de-anonymizing the relationship between the input and output addresses of mixing services [107].

Berdik *et al.* [108] have studied the role of BC in information systems as it becomes increasingly important when combining software and hardware utilities to develop an all-encompassing information system. They discussed how BC can be leveraged to develop or improve the next generation of information systems in modern enterprises, such as geographical information systems (GIS), healthcare information systems (HIS), media, energy, and financial information systems (FIS). Fig. 16 shows the percentage of BC implementation in various sectors, according to data reported by this reference.

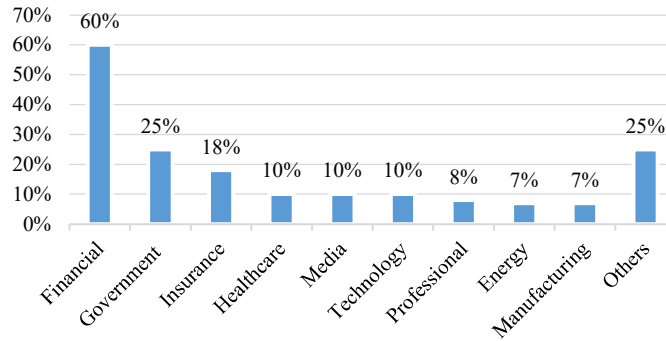

**Fig. 16:** The percentage of BC implementation in various sectors.

As demonstrated in Fig. 16, financial services, by far, have placed on the top BC representation. A decentralized architecture of BC enables a considerable deal of security and stability. However, the infancy of current BC-enabled systems has also caused some uncertainties about the capability of BC to appropriately manage data on a scale required by most businesses, in addition to *fault tolerance, storage,* and *broadening use*.

Blockchain technology serves as a foundational infrastructure for an array of secure services. This encompasses money transfer, supply chain management, and smart contracts. Each of these applications will be explored in the following three subsubsections.

#### 5.1.3.1 Money Transfer

Since its inception, the digital currency has taken the financial markets by storm. As a representative of digital current, Bitcoin is based on BC and has enabled a decentralized payment system based on peer-to-peer transactions. It has attracted the attention of worldwide businesses, consumers, and investors because of the uniqueness of its payment protocol. Bitcoin is proliferating so that major retailers, like AT&T, Overstock, and Microsoft, have adopted Bitcoin as a payment payload [109].

Du *et al.* [110] highlighted the strategic impacts of BC on organizations and discussed how to practically implement BC in organizations rather than studying technology's potential impacts. The researchers employed affordance-actualization (A-A) theory as the theoretical lens to conduct a case study of BC implementation in a real organization. They presented three affordances: (*i*) *subsidiaries and suppliers can settle payment directly*, (*ii*) *transaction participants can automate transactions*, (*iii*) *small suppliers can secure loans from financial institutions*, and a process model for actualizing these affordances. Their process model extended A-A theory by adding an experimentation phase where BC's payment system was added to the organization and tested through conceptual adaptation and constraint mitigation. Fig. 17 indicates the structure of the proposed BC-based payment system. As shown in Fig. 17, the proposed BC-enabled payment system consists of five main steps, from initiating a payment transaction to secure transfer of money between entities. This payment system can pave the way toward a secure money transfer mechanism in mobile payment platforms [111].

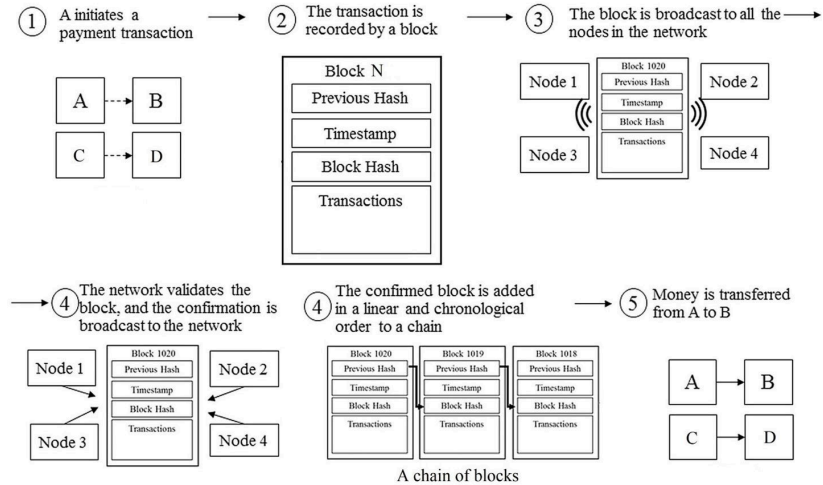

**Fig. 17:** The structure of a BC-powered payment system adopted from [110].

It is worth noting that while BC-enabled payment methods are significantly more secure than current systems, there is always a possibility of hacks due to the existing uncertainty of Bitcoin and also a delay in Ethereum markets. For instance, 1.1 million Bitcoins were hacked and robbed between 2013 and 2017. Considering the average price for a Bitcoin in that year (i.e. $7572 in 2018), an equivalent of 8.9 billion US dollars was lost, which emphasizes the importance of confronting these illegal actions [112].

Besides uncertainty, crypto-mining attacks have emerged as state-of-the-art web-centered attacks, which break the confidentiality, integrity, and availability of the victim's system. In this attack, the attacker abuses the victim's resources, including GPU and CPU, to mine and produce cryptocurrency. This attack will be completed in combination with other types of attacks, including social engineering and watering hole attacks, to deceive the victim into visiting infected web pages. This kind of attack caused a paradigm shift to web-centered attacks, which disregard the need for third-party exploiters. To encounter these attacks, Zimba *et al.* [113] have explored Cryptojacking injection attacks (a state-of-the-art web-centric attack vector) in the crypto-mining attacks landscape. This research also proposed a formal manner based on finite state machines (FSM) to model such attacks. Their proposed model indicates how attackers can infiltrate and compromise core e-commerce components to use their resources to generate Monero cryptocurrency.

In another recent study, Hassani *et al.* [114] reviewed the effects of BCs in the financial and banking sector, as well as the challenges of big data facing BC infrastructure. Although they found solid evidence of adopting BC technology in a number of banks, their findings further indicated a severe need for conducting more research to investigate deployment aspects of banking with BC. Their study also highlighted the necessity to tackle the barriers currently hindering the adoption of BC in banking across the globe. In conclusion, new BC technology is increasing its significance as a solution in dealing with cybersecurity threats and promising a more secure trend for the evolution of the financial sector.

### 5.1.3.2 Supply Chain Management

In Industry 4.0 era, the importance of sustainability in supply chains (SC), as a critical component in today's digitalized circular economy, is more evident than ever [115]. Supply chain finance (SCF) is crucial in facilitating the efficient operation of corporate activities within the contemporary global supply chain. In 2020, the global market for SCF reached a noteworthy milestone, with a valuation of 46 billion USD. Currently, the global market for BC-enabled SCF is valued at USD 84,540 million and is projected to expand at a CAGR of 33.6% from 2021 to 2026[13]. BC also plays a pivotal role in building trust in the system by employing a decentralized 'trustless' network, which is one of the vital factors influencing BC adoption [116].

The study by Paul *et al.* [117] examined the effects of the COVID-19 pandemic declaration on the market value and trading volume of SCF businesses, focusing on banking. The researchers employed an event study methodology to examine the impact of a specific event on the valuation and trading volume of firms in the sustainable and socially responsible investment sector. Their analysis revealed a substantial decrease in valuation and an increase in trading volume for these firms following the event. On the other hand, organizations that have implemented BC-enabled SCF systems are shielded against potential losses in valuation and the associated fluctuations in trading, as well as security threats. They also examined whether firm-level heterogeneity affects value protection or loss owing to the announcement using cross-sectional regression analysis. It has been shown in this study that increased investment in research and development of BC-enabled SC (being part of a blockchain consortium rather than working alone) serves as effective measures in mitigating threats and losses.

---

[13] https://www.marketwatch.com/press-release/blockchain-supply-chain-finance-market-size-2021industry-analysis-size-share-trends-marketdemand-growth-opportunities-and-forecast-2026-2020-12-09.

In another recent study, Wamba *et al.* [118] systematically investigated and evaluated the feasibility of replacing the current concept of money in new supply chains. They reflected the pros and cons of such replacement in several industries, aiming to address the existing challenges in FinTech. For the possibility of money replacement, this research has assessed the evolution of Bitcoin, Blockchain, and FinTech over time. The outcomes and findings of this study demonstrated that these technologies are drastically being persuaded, and sooner or later, this replacement will take place. Hence, businesses and organizations should embrace them as a competitive advantage. Furthermore, organizations need to conduct more research on these technologies to have a better understanding. With this knowledge and novel technology, they can optimize their current strategies and policies and develop insights for decision-making.

Whereas the adoption of BC in SC has offered many facilities, it has increased the number of cyber threats and the system attack's surface, such as business data leakage, intellectual property theft, and interruption of business operations. Therefore, BC-enabled SCs have some risks and uncertainties as their resilience from a research perspective is in an early stage, and supply chain networks need several cybersecurity considerations and threat management policies [119]. To further recognize cyber threats in SC, Syed *et al.* [120] studied existing risks and cyber threats against supply chain systems. Using a systematic asset-centric threat modeling method, they included more than a hundred relations between threats, assets, and countermeasures related to SC traceability. They also adopted the STRIDE threat model to describe common threats. STRIDE includes spoofing, tampering, repudiation, information disclosure, denial of service, and elevation of privilege. This model consists of a multilayer architecture for SCs where the analysts are able to accommodate a concrete implementation and perform secure traceability and falsifiability by providing the sources used to establish the relation, i.e. threats, assets, and countermeasures.

### 5.1.3.3 Smart Contracts

Smart contracts leverage BC technology to verify that two parties have signed an agreement, as well as to ensure transactional integrity [108]. A smart contract is a piece of code aimed to enforce the execution of a deal between two or multiple parties. The program code can be self-executed and deployed as BC infrastructure transactions; hence, no modification in its code is possible. This feature significantly reduces the likelihood of fraud and forgeries. Smart contracts also offer many advantages, including removing/minimizing the possibility of money laundering and supplying terrorist groups, compared to the current trading systems. Meanwhile, the rapid deployment of smart contracts in businesses raises a need for security verifications since failures in smart contracts or any security breach can bring about substantial financial loss and global panic [121]. To this aim, Almakhour *et al.* [122] have proposed a framework for formal verification of composite smart contracts security using FSM. They provided two verification types, one suitable for all smart contracts with standard properties, and the other one considers the context-dependent properties, called 'specific properties' varying based on the domain of usage, focusing on the financial sector. A formal security verification can make use of smart contracts possible in SC to replace traditional trades.

## 5.2 Human-originated Approaches

The second subclass of defense strategies focuses on the human element, addressing behaviors and activities. These strategies, which primarily hinge on cultural education and technical training, aim to mitigate cybersecurity knowledge gaps. By doing so, they hold the potential to neutralize a significant number of security threats targeting FinTech systems and their clients.

### *5.2.1 Cyber-culture and Security Awareness*

Cyber-culture and security awareness is recognized as the most important missing part in FinTech services, which can address a wide range of cyber threats, including insider attacks, social engineering, data leaks, and Internet frauds.

In this regard, Varga *et al.* [30] presented a method to evaluate the properties of a Common Operational Picture (COP) as a base for establishing a Cyber Situational Awareness (CSA) team for financial sector actors. In this study, a dataset was collected through surveys and interviews with pivotal players in the financial sector concerning national-level crisis management. The questionnaire designed by the authors included ten questions regarding the orientation category of cyber threats. The main questions focused on five W questions (Ws) and involved the information elements required in a COP and the perception of cyber threats. The number of responses received was 42 out of 70 participants. Interviews were conducted with 270 persons representing substantial Swedish financial sectors, including security dealers, insurance companies, banks, and central financial system players like stock market holders. This study classified sources of cybersecurity operational risks into four classes, i.e. failed internal processes, systems and technology failures, actions of people, and external events. Their analysis results demonstrate that social engineering is the most threatening attack vector. Insiders and the lack of real-time intrusion detection were also recognized as potential threats. Moreover, their reports indicated that the most common technology-based cyber-attacks are DoS and malware attacks - Dridex, Shylock, and Zbot - by 55% and 52% of the respondents, respectively. The authors also recommended seven provisions for having a complete CSA for cyber defense against threats in FinTech. They included the awareness of the current situation, why and how the current situation began, how the situation evolved, the adversary of behavior, the impact of attacks, the integrity and quality of the situational awareness information, and assessment of probable future circumstances against the current state.

One of the important limitations of the recent reference is the validity of the conclusions that might have been obtained under particular situations in Sweden and cannot necessarily exist elsewhere. Besides, the Swedish financial sector comprises a few actors and the number of participants in the questionnaire seems small.

Dealing with social engineering attacks and Internet fraud severely requires a systematic analysis of possible attack patterns to accurately detect the initial steps by which an attack gets started. Detecting these attacks from the initializing steps makes countermeasures much more effective. This highly needs an appropriate knowledge of cyber-culture and security awareness. Airehrour *et al.* [81] proposed a model to identify the initial steps of social engineering attacks. The procedure of this model for a single-stage social engineering attack is illustrated in Fig. 18. As indicated in this figure, the model has consisted of five steps, i.e. reconnaissance, planning and preparing, launching a phishing attack, information capturing, and finally, exploiting the collected information, in which a defensive action has been offered for each step of the attack. This model can be extended and applied to multi-stage attacks since steps in the multi-stage attack are also captured in a single-stage attack scenario.

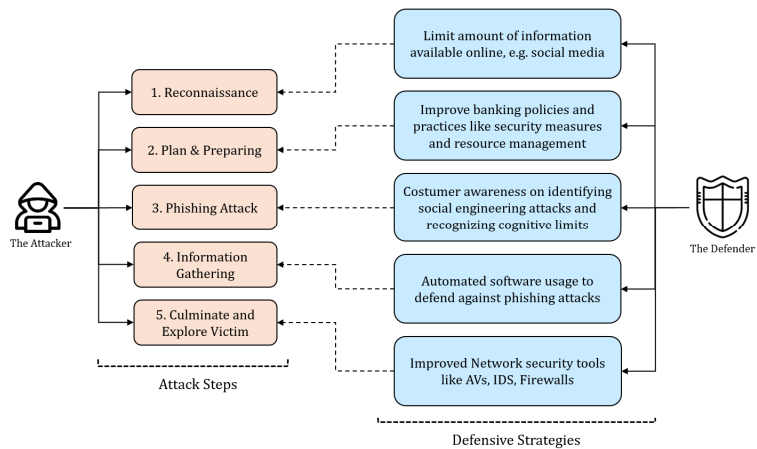

**Fig. 18:** The procedure of the user-reflective model to confront social engineering attacks, adopted from [81].

*5.2.2 Cybersecurity Training*

The current worldwide workforce's cybersecurity skills have faced a significant critical level of shortage; this issue has aggravated in the financial sector [123]. Meanwhile, training well-skilled cybersecurity staff in the numbers required by the financial sector is a time-consuming task. It needs plenty of related experience and practical efforts. However, employing hands-on exercises is one of the best practices for quickly learning cybersecurity skills. Leon *et al.* [123] have designed a tutorial-based and classroom approach to cybersecurity instruction, presenting six hands-on tutorials on operating systems security.

Cybersecurity education, awareness raising, and training (CEAT) undeniably affect the vitality of Internet services for various domains at the national and international levels. CEAT includes one out of five aspects of a larger cybersecurity capacity-building model (CMM) developed by the Global Cybersecurity Capacity Centre. This aspect and its related indicators of education, awareness, and training in cybersecurity were presented by Shillair *et al.* [124]. This study included a multinational analysis of CEAT outcomes on data gathered from 80 countries considering different parameters, such as the Internet usage scale and the country's wealth. The results indicated that CEAT could considerably and positively impact the vitality of Internet usage for enabling financial services. Besides, the distribution of CEAT scores recognized some key factors to determine the level of maturity for low-income and under-developing countries. This study also reported the responses from these countries as the basis for offering recommendations for making policies and practices to meet the need for actual awareness, education, and training.

Some organizations have run several awareness and training courses to improve cybersecurity knowledge, while fewer have considered cybersecurity knowledge sharing. To fill this gap, Alahmari *et al.* [125] produced a bespoke software application. This application has developed a game to deliver security training based on the theory of a transactive memory system (TMS). In addition, the authors have performed an empirical study on one hundred organizations in Saudi Arabia to assess the impacts of this training application on the employees. Their evaluation results indicated that their TMS-based training application significantly has increased cybersecurity knowledge and awareness in those organizations.

### *5.2.3 Cyber-threat Intelligence Sharing*

Cyber-threat Intelligence (CTI) sharing has proved to be an effective solution to detect suspicious traffic from notorious cyber actors; however, there are two main challenges concerning CTI sharing in FinTech. First, using low-level Indicators of Compromise (IOCs), including IP/Port addresses and domain names to attribute cyber-attacks, are highly prone to be deceived by attackers through changing, spoofing, proxies, or anonymizing. These deceiving approaches have widely been employed in novel attack patterns, particularly APTs, mainly when there would be financial gains for attackers. Second, current methods are usually based on manual analysis of IOCs taken from security information and event management (SIEM) systems, firewalls, and IDS applications. Manual analysis of IOCs is extraordinarily time-consuming and tiresome, especially when a large volume of data is needed for a thorough inspection. Hence, several studies have introduced automated approaches using machine learning techniques.

To this aim, Noor *et al.* [24] disclosed six cyber-threat actors (CTAs) and presented an automatic framework to derive high-level IOCs from existing unstructured human-readable CTI reports applying natural language processing (NLP) and then attribute them. Software tools, malware pieces, and tactics, techniques, and procedures (TTP) used in attack patterns of cyber kill chains are considered high-level IOCs. This study collected 327 CTI documents from 26 sources between 2012 and 2018 to attribute 36 threats. They have proposed a vocabulary to map the attack pattern queries to conceptual meaning by labeling high-level IOCs from the Adversarial Tactics, Techniques, and Common Knowledge (ATT&CK) dataset[14]. ATT&CK taxonomy includes 188 TTP and 146 malware programs commonly utilized by CTAs. Then, using Latent Semantic Analysis (LSA) based on MITRE taxonomy[3], the records were indexed, and a semantic search was conducted to seek attack patterns. Five machine learning and deep learning algorithms were employed to train the detector model. The authors used 10-fold cross-validation to evaluate the performance of their proposed model. The results indicated that the authors' automated semantic system can detect TTPs more accurately than manual approaches. The output's figures also demonstrated that the deep neural network (DNN) attributes cybersecurity threats with higher precision, i.e. 94%, compared to other classifiers.

In another effort, Cascavilla *et al.* [126] performed a Topic Modeling Analysis (TMA) to extract features to recognize CTIs from the Dark- and Deep-Web. This research further presented an overview of features and risk parameters to detect threats of such cybercrime activities. It also provided a taxonomy of threat engineering and management to defeat their actors. Ultimately, this research concluded with some recommendations, including website features and the degree of anonymity for Law Enforcement Agencies to confront cyber threats, particularly threats against the financial sector.

To hasten the international automated CTI sharing with a focus on IP addresses in business-to-business trade, Sullivan and Burger [127] published their findings of a cross-discipline research project on perusing the data of 34 jurisdictions in the European Union (EU). Despite the previous CTI sharing reviews developed at regional or country levels, this study focused on a global scope for CTI sharing. This study underscored the importance of recognizing whether an IP address can be shared with third parties as a cyber threat intelligence legally or not. After reviewing the EU General Data Protection Regulation (GDPR) agreed upon and officially adopted by the EU Parliament in 2016 and US Data Protection Directive 95/46/EC (1995 Directive), they concluded that automated business-to-business (B2B) CTI sharing of that data could be carried out in the public interest under Article 6(1)(e) of the GDPR and its equivalent in the 1995 Directive. This study proved that threat intelligence sharing is in the public interest, and this interest can override the individual rights of a data subject under Article 8(1) of the Charter of Fundamental Rights of the EU and its equivalent in the 1995 Directive. However, they highlighted some deficiencies with Articles 13 and 14 of the GDPR concerning notification of the data subject and recognizing bad/suspected actors. Hence, they proposed some policies and procedures to bridge these gaps and pave the way for global automated CTI sharing in a lawful manner. The authors attempted to convince the European Council and Parliament to adopt their proposal.

---

[14] https://attack.mitre.org/

It is crystal clear that legal shortcomings exist in the legislation for CTI sharing in FinTech, which has negatively affected the effectiveness of this solution; hence, there is an urgent need to update/introduce new legislation, particularly regarding automated sharing.

## 5.3 Procedure-related Approaches

In concluding our review of defensive solutions, we outline methodologies put forth by five papers classified within the procedure-related category of defenses. These solutions are designed to identify and close the gaps in both internal and external policies and procedures of entities operating within the FinTech sector.

### *5.3.1 Risk Management Policies*

Risk management is the process of implementing an effective cybersecurity plan, which has an undeniable role in international regulations for financial data protection. Some studies have reflected that a complete perception of cyber risks is needed to ensure sufficient security procedures in institutes and banks. The accuracy of risk management highly depends on our knowledge of companies' assets and features, especially in the financial sector, since the main assets are money, securities, bonds, and stocks. Hence, a digital-asset assessment procedure is desperately needed at the beginning step to allow suitable protection against cyber threats, particularly for FinTech applications, in which various technical and commercial assets have enclosed companies [128]. Unfortunately, proven models for measuring cyber risks are scarce, and a few existing models have some limitations, such as a lack of generalization, since most consider merely analyzing past data to extract statistical and probabilistic models. Besides, quantitative approaches such as HTMA [129] and FAIR [130] are based on the likelihood of subjective evaluation of events, where the lack of sufficient quantitative data led to inaccuracy with these models. On the other hand, while a sheer amount of quantitative data allows a reliable assessment, it makes the analysis of data usually time-consuming and sometimes not feasible [131]. Therefore, providing reliable models for cyber risk exposure remains an open challenge. Besides, the financial sector needs a fully customized risk management policy for cyber-insurance since the assets in this sector are invaluable, easy to steal, and different from other domains.

To bridge this lack, Santini *et al.* [23] suggest combining objective data and probabilistic approaches as input for the HTMA model to evaluate the cyber risk exposure in institutes and enterprises. The authors presented several data-driven key risk indicators (KRIs) to be employed in the HTMA model in order to decrease the subjectivity margin since lots of subjectivity in the risk assessment might undermine the trustworthiness of results. They have combined a data-driven approach with a quantitative approach to assess the likely risks of economic loss in a corporation and score its cyber exposure. The CIS 20 critical security controls were used to indicate the effectiveness of applied measures. They have claimed that their proposed approach can defeat the limitations of current risk assessment models based on estimating the occurrence likelihood and the observation of incidents that have taken place in the past. Also, it has been claimed that this solution is practically applicable to the financial sector and industrial manufacturing world, which is a promising point.

### *5.3.2 Cyber-insurance*

Insurance significantly decreases financial risks and allows protection against unexpected financial loss; however, the lack of sufficient regulations has led to imperfect coverage of cyber incidents and cyber-attacks. Pisoni *et al.,* in their study [132], indicated the requirements and specifications for these systems from a regulation standpoint. Then, inspiring Tech-aided advisory systems in all financial services, e.g. the AI-powered advisor that helps clients with loans, they proposed a novel AI-based approach in which recommender techniques were used to develop such an advisor to suggest insurance coverage. This approach also included a transparent and trustworthy system design responsible for designating which data can be relied upon for the financial players. Insurance companies can deploy this system to ensure the completeness of cyber insurance policies and cyber-attack coverage.

### 5.3.3 Regularization Technology

Regularization in FinTech is one of the primary requirements, and in some cases like cryptocurrencies, is a missing point. Regulatory technology (RegTech) employs AI-enabled techniques for intelligent identification and early risk warning, which can be a powerful tool to reach financial regulation. Chao *et al.* [133] have reviewed intelligent technology applications in financial stability regulation to analyze and highlight the limitations of RegTech. To achieve financial stability, this research implemented RegTech by forming a framework for the applications of machine learning, deep learning, knowledge graphs, complex networks, and dynamic systems.

In another study, Currie *et al.* [134] conducted a longitudinal study on the UK financial services industry. They investigated empirical data gathered over 12 years examining the deployment of an investment management system in 8 financial firms. This study proved how RegTech can be used in financial services to mitigate pre- and post-crisis, including cyber threats. Their findings indicated that whereas a 'tsunami' of financial regulations was introduced by government institutions and private investors in response to the financial crisis so far, regulatory bodies should revise and impose compliance mandates on financial firms to force them to adapt their financial technologies in an ever-changing multi-jurisdictional regulatory landscape.

## 6. Key Findings

In this section, we discuss the results of our survey in detail. We address the research questions outlined at the outset, share insights gained, and offer recommendations for countering cyber threats in the FinTech sector. We eventually suggest future research directions. Acronyms used in this section can be found in Table 4, where they are expanded.

**Table 4:** The list of acronyms used in this section.

| Acronym | Expanded from | Acronym | Expanded from |
| ---: | --- | ---: | --- |
| A | Application | I | Infrastructure |
| AC | Access Control | I Fraud | Internet Fraud |
| And | Android | Int Det | Intrusion Detection |
| Att | Attacks | Imp | Implementation |
| Awar | Awareness | Malw | Malware |
| BC Inf | Blockchain Infrastructure | Mng | Management |
| CCS | Cyber-culture & Security | Mn Tns | Money Transfer |
| Chn | Channel | P | Platform |
| C Insur | Cyber Insurance | PW | Power |
| CS | Cybersecurity | Ref | Reference |
| DB | Data Breach | SC | Supply Chains |
| Digi Extr | Digital Extortion | Sim | Simulation |
| DL | Data Leak | Smrt Ctr | Smart Contract |
| Env | Environment | Vulnr | Vulnerability |
| Esp | Espionage | Win | Windows |

### 6.1 Most Recent Threats (RQ1)

To address RQ1, threats from various natures and origins, i.e. technology-related, human factors, and procedures, can cause severe hazards and put FinTech into jeopardy. According to our survey, technology-related threats with six attack patterns are the most potent type of threat against FinTech; in contrast, procedures with only two attack vectors are the least possible threat, while threats that originate from human factors are in an intermediate state. Fig. 19 displays the distribution of these threats and their corresponding defenses, organized by the percentage of related publications, and sorts them based on the percentage of published papers in each category. This percentage was derived from the number of papers retrieved using the query described for this survey, adhering to the eligibility criteria explained in Section 3.

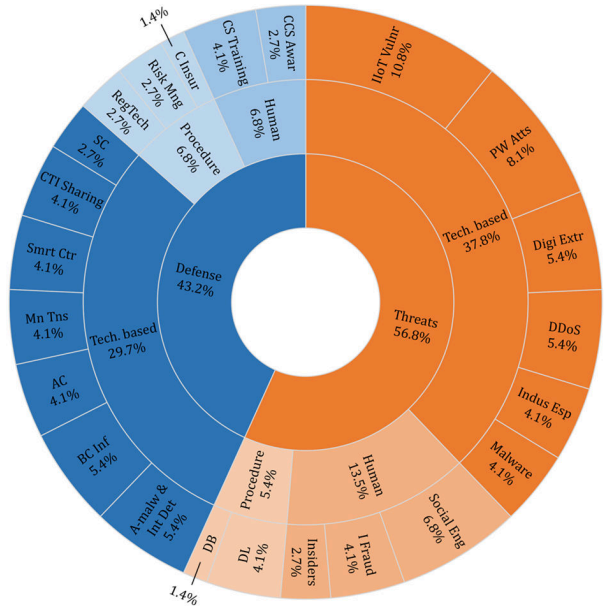

**Fig. 19:** The percentage of published papers in each category of threats and defenses with a color spectrum.

## 6.2 Most Destructive Threats (RQ2)

In addressing RQ2, we ranked threats based on their destructiveness across various sectors, emphasizing the financial sector, as shown in Fig. 20. In this figure, sectors that are highlighted have experienced more frequent attacks and sustained greater damages from the corresponding threat(s). In the financial sector, Malware Attacks emerge as the most prevalent and destructive, followed by DDoS and Social Engineering Attacks. This threat distribution is distinct for other sectors, e.g. Internet Fraud and Industrial Espionage rank as the most common threats in the health sector.

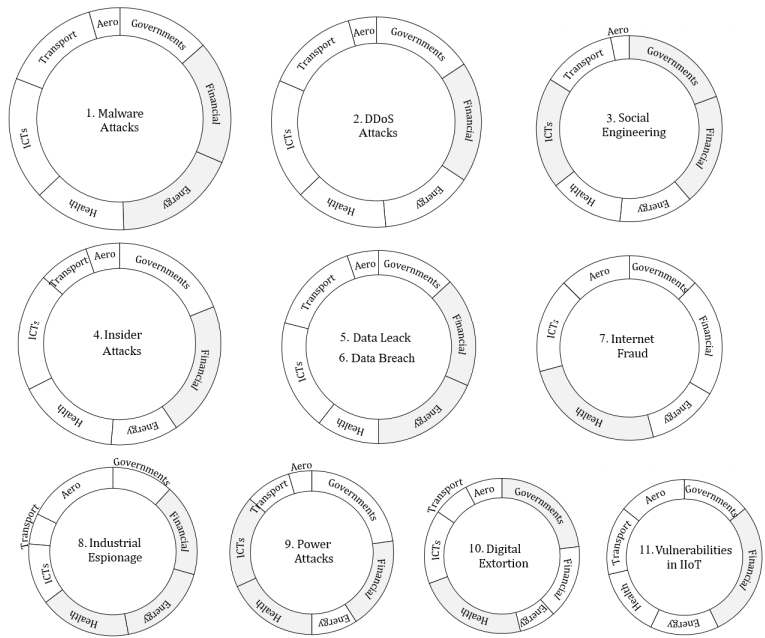

**Fig. 20:** The distribution of cyber threats against various sectors, where a bigger circle indicates a more destructive threat.

## 6.3 Existing Defense Strategies (RQ3)

In addressing RQ3, we present a comprehensive list of defensive solutions in Table 5, detailing the threats each solution can mitigate. The table offers a comparative analysis of these solutions from multiple perspectives: their primary concept and contribution, their domain and category, the specific service layer and channel they are designed for, and finally, the environment in which the effectiveness of the solution was evaluated.

**Table 5:** A list of the best-offered defensive solutions to mitigate cyber threats against FinTech.

| Reference | Main idea and contribution | Domain | Category | Mitigated Threats | | | | | | | | Service Layer | | | Chn | Env |
|---|---|---|---|---|---|---|---|---|---|---|---|---|---|---|---|---|
| | | | | Malw | DDoS | Digital Extortion | Insiders | Social Eng. | Internet Fraud | Data Breach | Data Leak | I | P | A | | |
| Noor et al. [24] | An automatic framework to derive high-level IOCs from existing unstructured human-readable CTI reports applying Natural Language Processing (NLP) | Technology | CTI Sharing | | ✓ | ✓ | | | | | | | | ✓ | Web | Imp |
| Cascavilla et al. [126] | A Topic Modeling Analysis (TMA) to recognize the most dangerous cyber threats in the Dark-Web and Deep-Web. | | | | | | ✓ | ✓ | | | | | | ✓ | Web | Sim |
| Sullivan and Burger [127] | Investigating the legal aspects of automated CTI sharing with a focus on IP addresses for international B2B players by perusing 34 jurisdictions' data. Recommendations to revise the current GDPR in the EU and the 1995 Directive in the US. | | | | | | ✓ | ✓ | ✓ | | | | | ✓ | | |
| Sherman et al. [39] | A semi-supervised framework using ML-based algorithms, including SVM and RF, for the detection of Ransomware attacks | | Anti-malware and Intrusion Detection | ✓ | | ✓ | | | | | | | ✓ | | Win | Imp |
| Min et al. [41] | Proposed a defensive approach against Ransomware attacks, containing a detection model and a recovery method based on Solid-State Drive backup. | | | ✓ | | ✓ | | | | | | | ✓ | | Win | Imp |
| Bai et al. [96] | Using a triadic graph mining approach to detect Android Banking Trojans based on APIs. | | | ✓ | | | | | | | | | ✓ | | And | Imp |
| Dassouki et al. [97] | Proposing an intrusion detection system to deal with DDoS attacks, including SIP flooding and TDoS attacks, against authentication mechanisms in mobile/telephone banking systems. | | | | ✓ | | | | | | | | ✓ | | Multi | Imp |
| Malaquias and Hwang [98] | Conducting an empirical study on trust assessment in mobile banking systems through Structural Equation Modeling and Confirmatory Factor Analysis on 1077 questionnaires collected in Brazil in 2016. Indicating barriers to building trust in FinTech systems and accelerating its adoption in banks. | | Access Control | | | | ✓ | | | ✓ | ✓ | | | ✓ | | |
| Hernandez et al. [99] | Present S&P solutions for data protection on FinTech platforms to build trust in financial services. | | | | | | ✓ | | | ✓ | ✓ | | | ✓ | | |
| Qiu et al. [100] | An attribute-based access control system leveraging a proactive user-centric data security approach to be employed in the financial sector. | | | | | | ✓ | | | ✓ | ✓ | | ✓ | ✓ | Multi | Imp |
| Lei et al. [103] | Profiling 443 articles to present a new classification of BC applications in information systems and highlighting the promising role of BC in the future systems designs of financial services. | | BC Technology | | | | ✓ | | ✓ | | ✓ | | | ✓ | Multi | |
| Raddatz et al. [105] | Using BC as a data store infrastructure in financial-related applications instead of traditional database systems. Promoting data privacy and transaction integrity leveraging BC. | | | | | | ✓ | | ✓ | ✓ | ✓ | ✓ | | | | Imp |
| Berdik et al. [108] | Studying the role of BC in information systems. Indicating how BC can be leveraged to develop or improve the next generation of information systems in modern enterprises, focusing on financial information systems (FIS). | | | | | | ✓ | | ✓ | | ✓ | | ✓ | | Muti | Imp |
| Zimba et al. [113] | Exploring cryptojacking injection attacks in the crypto-mining attacks landscape and studying a state-of-the-art web-centric attack vector. | | Smart Contracts | | | ✓ | ✓ | | ✓ | | ✓ | | ✓ | ✓ | Web | |

| Reference | Description | Category | Subcategory | C1 | C2 | C3 | C4 | C5 | C6 | C7 | C8 | C9 | C10 | C11 |
|---|---|---|---|---|---|---|---|---|---|---|---|---|---|---|
| Hassani et al. [114] | Reviewing the effects of implementing BC in the financial and banking sector, as well as the challenges of big data facing BC infrastructure. Reporting solid evidence of adopting BC technology in several banks. | | | | | | ✓ | | | ✓ | | ✓ | | Web | Imp |
| Sindhwani et al. [115] | Analyzing the adoption of BC in supply chains and its effects, including the integrity and sustainability of product manifest, data sharing, traceability, and end-to-end visibility of products in Industry 4.0. | | | | | | ✓ | | | ✓ | | | ✓ | | |
| Wamba et al. [118] | Investigating the SC industry to determine the pros and cons of employing SC instead of money in various industries to deal with challenges in FinTech. | | Money Transfer | | | | ✓ | | | ✓ | | ✓ | ✓ | | |
| Syed et al. [120] | Studying existing risks and cyber threats against supply chain using a systematic asset-centric threat modeling method. Adopting the STRIDE threat model to describe common threats in supply chains and accommodate a platform for secure traceability and falsifiability. | | | | | | ✓ | | | | ✓ | | ✓ | Web | Imp |
| Suegami [121] | Proposing a cryptography-based obfuscation technique to evaluate the trust level in smart contracts and cryptocurrencies. | | Supply Chain Management | | | | | | | ✓ | | ✓ | ✓ | Web | Imp |
| Almakhour et al. [122] | A new method to perform formal verification in smart contracts employing finite state machine. | | | | | | | ✓ | | ✓ | | | ✓ | Web | Imp |
| Varga et al. [30] | Presenting a Common Operational Picture (COP) as a base for establishing Cyber Situational Awareness (CSA). | | Cyber-culture and Security Awareness | | | | ✓ | ✓ | ✓ | | | | | | |
| Airehrour et al. [81]* | Suggesting a User-reflective model for recognizing and mitigating social engineering attacks in Banks of New Zealand. | Human | | | | | | ✓ | | | | | | | |
| de Leon et al. [123] | Proposing a tutorial for implementing practical security measurements at the OS level. | | | | | | ✓ | | | | ✓ | | | N/A | N/A |
| Shillair et al. [124] | Providing some proposals for cybersecurity training and awareness using pieces of evidence at the national level in 80 countries worldwide. | | Cybersecurity Training | | | | ✓ | ✓ | | ✓ | ✓ | | | | |
| Alahmari et al. [125] | A new model and mobile applications for cybersecurity training, awareness, and knowledge sharing. | | | | | | ✓ | ✓ | | ✓ | ✓ | | | | |
| Santini et al. [23] | Combining objective data and probabilistic approaches as input for the HTMA model to evaluate the cyber risk exposure in institutes and enterprises. | | Risk Management | | | | ✓ | | | | | | | | |
| Masoud et al. [131] | A framework to determine the empirical indicators for assessing cybersecurity risks in the financial sector based on financial reports. | | | | | | ✓ | | | ✓ | ✓ | | | | |
| Pisoni et al. [132] | Determining the requirements and specifications for cybersecurity insurance from a legal standpoint. Developing an AI-powered recommender system to suggest insurance coverage for financial services. | Procedure | Cyber-insurance | ✓ | ✓ | ✓ | | | | | | | N/A | N/A |
| Chao et al. [133] | Analyzing AI applications in financial stability regulation to overcome the limitations of RegTech to reach financial stability. Proposing a RegTech framework for machine learning, knowledge graphs, complex networks, and dynamic systems applications. | | RegTech | | | | ✓ | | | | | | | | |
| Currie et al. [134] | Analyzing empirical data gathered over 12 years examining the deployment of an investment management system in 8 financial firms in the UK from a regulatory landscape. Proving how RegTech can be used in financial services to mitigate pre- and post-crisis, including cyber threats. | | | | | | ✓ | | | | | | | | |

As shown in Table 5, prominent features and limitations of the offered defense methods were indicated. From one comparison, we determined that using standard datasets is a necessary and inevitable factor in performing a valid assessment of cyber threats. Hence, it is inadvisable to place complete trust in studies that base their findings on experimental datasets, especially when dealing with critical infrastructures like financial technology. Additionally, some studies have overlooked the assessment of time complexity for their proposed feature selection algorithms. An effective and on-time response to cyber-attacks is heavily influenced by the computational complexity of these algorithms.

## 6.4 Applying Defenses-to-Threats (RQ4)

This subsection addresses RQ4. Fig. 21 presents a taxonomy for cybersecurity threats discussed in Section 4. Additionally, it provides a hierarchical structure of defensive strategies derived from the papers reviewed in Section 5. Besides, this subsection details the application of these defensive solutions in countering imminent threats.

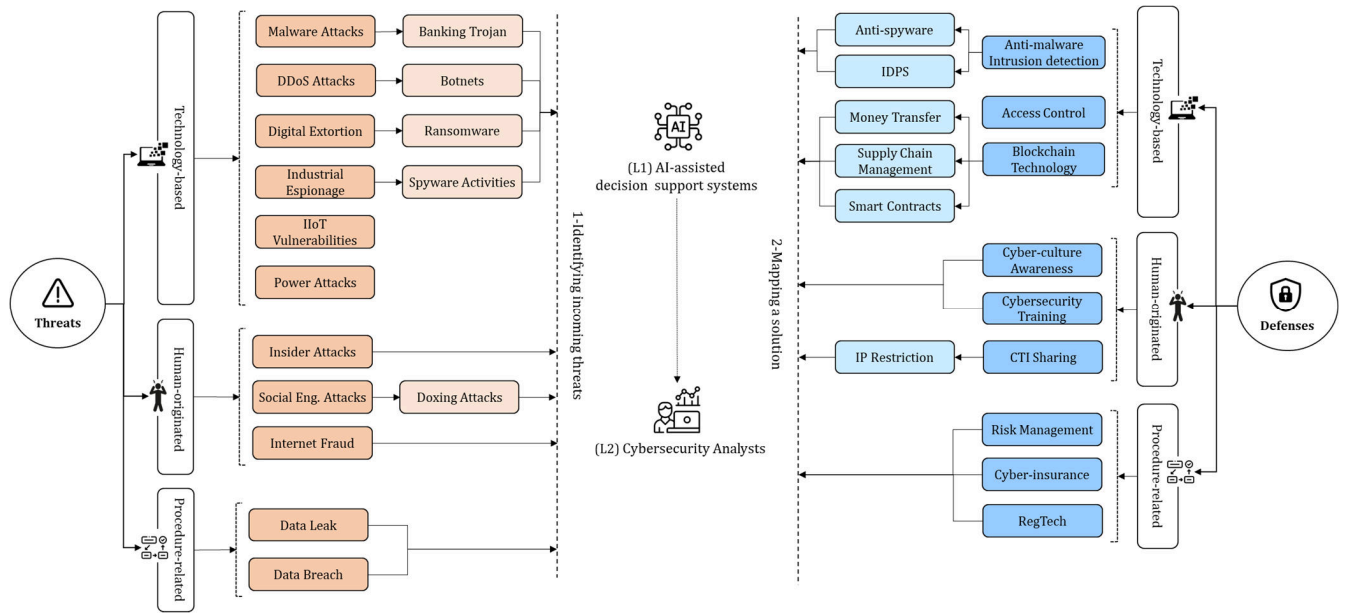

**Fig. 21:** A taxonomy of defensive strategies versus a hierarchy of cybersecurity threats in FinTech.

As shown in Fig. 21, threats originated from three main sources, i.e. technology-based, human-originated, and procedure-related. Mirroring this structure, defensive solutions are similarly categorized. Each defensive method is capable of mitigating a single or multiple threats. Two strategies exist for applying these solutions, represented as defensive layers, L1 and L2. The first utilizes AI-assisted decision support systems for automatic threat response, while the second requires manual intervention by a cybersecurity analyst.

L1 and L2 are responsible for identifying incoming threats and selecting appropriate defensive strategies from available solutions. This vital task is carried out in the first layer using *AI-assisted decision support systems*. In the case of any failure or an unprecedented situation in which the first layer (L1) would not be able to complete the task, and the second layer (L2), i.e. *Cybersecurity analysts*, will take control to recognize the type of incoming attack and make a suitable decision to pick the best possible defensive solution. Additionally, the cybersecurity analyst can surveil and supervise L1 procedures and decisions. However, this approach has a significant limitation since human analysts cannot handle all the incoming threats, considering the sheer amount of today's cybersecurity attacks against FinTech. This underscores the indispensable need for an AI-assisted decision support system.

## 6.5 Lessons Learned and Recommendations (RQ5)

Lessons learned from studying identified papers regarding cyber-attacks that occurred in FinTech as well as our recommendations, are reported in Table 6. These recommendations can be beneficial for businesses, banks, institutes, and entities active in the financial sector to bear ever-growing cybersecurity threats, take an effective response, and last their businesses.

**Table 6:** Lessons learned from previous cyber-attacks and our recommendation to address them.

| Topic | Lessons learned | Our recommendations |
|---|---|---|
| All human-originated threats | - Compared to well-developed intelligent defensive systems, human beings are the weakest link in the digital system's life cycle, where social, psychological, and cultural factors can influence their behavior. In this case, the human-originated factors are sometimes the least challenging and most accessible possible gateway for infiltrating secure facilities.<br>- Technology is being used to deceive humans as the most recent trend of cyber threat in FinTech. | - Cyber-culture and security awareness to negate a large number of threats, including both technology-based and human-originated threats from the very beginning. |
| Insider Attacks | - As a result of an increase in the number of insider attacks, recent solutions and security debates should be enlarged to include trusted employees, contractors, and business partners with more focus on insider threats [64].<br><br>- Trust is a vital factor for FinTech systems since it not only brings efficiency to financial institutions and banks but also can potentially improve clients' quality of life [98]. | - Providing pre- para- and post-employment policies and agreements to maintain the confidentiality of data and information at an appropriate level and guarantee it for a suitable period after the employment term.<br>- Post-employment has been the most neglected policy in many big companies and businesses<br>- Although BYOD policies are critical to employees and have an undeniable role in their interaction, they cannot be followed in the FinTech sector as they might be a potential source for the propagation of security flaws and vulnerabilities and, thus, loss of clients' trust. |
| Malware Attacks | - There is a hide-and-seek relation between security threats categorized in Section 4. For instance, digital extortion, DDoS, and industrial espionage are frequently carried out by malware programs. Therefore, in addition to direct attacks, malware programs can perform the role of attack vectors for other types of security threats. [51], [52]<br>- If the malware is launched and the attack gets started, much more effort will be needed to defuse the attack. It mainly needs stopping and restarting the systems, which might cause service interruption. These interruptions are extremely costly for the financial sector. | - Understanding malware's internal procedures and attack patterns is the most effective way to confront a broad range of cyber-attacks since the primary source, malware executable file, will be eradicated.<br>- Using Anti-malware applications that are tailored to work based on a hybrid malware analysis method, as they can simultaneously utilize the advantages of static and dynamic analysis methods.<br>- Being equipped with monitoring/intercepting facilities based on transparent hooks at kernel-level. IRP hooks or hooks with more depth in the OS kernel space are much more effective in detecting APT attacks of banking malware than other types of hooks [48].<br>- IRP hooks can be implanted using kernel-level filter drivers in different versions of Windows NT. |
| Banking Malware | - In marked contrast to a drastic incensement in the number of malware in recent years, according to credible resources, the number of banking malware, responsible for most catastrophic damage in the financial sector, has significantly decreased. Our study discovered and disclosed that they have not reduced in reality; they have become stealthier and more obfuscated, so they are detected less by AVs.<br>- Like the Spyware class, banking malware is one of the most challenging classes to detect by AVs as they are well-equipped with several obfuscation and metamorphism engines [135].<br>- Hardware trojans capable of stealing sensitive financial information might be embedded into a variety of equipment used in the FinTech sector that are not observable by visual inspection [26]. | - Using specialized Anti-spyware applications besides generic AVs for systems and servers that host financial services and FinTech applications to detect stealth banking malware.<br>- Generic AVs usually ignore scanning inter-process communication (IPC) for child processes/threads with a depth of three or more since it consumes plenty of system resources, resulting in system lags.<br>- Anti-spyware applications are tailored to intercept IPC communication with more depth than generic AVs - usually up to the depth of 7 parent/child processes or threads rather than 3 inter-app [136].<br>- Optical inspection or Scanning Electron Microscopy is highly recommended to detect hidden hardware trojans and security assurance of critical equipment employed by FinTech service providers [137]. |
| Digital Extortion | - There is always a significant chance of ransomware attacks for digital extortion from organizations, particularly those that are active in the financial sector. This chance was estimated to be 66% by Sophos[15] report in 2022.<br>- Over 90% of malware programs are tailored to run on the Windows OS [48]. | - Getting backups from the sensitive data and storing them on external and firewalled/air-gapped servers, aiming to prevent the propagation of ransomware to the backup servers.<br>- Employing different OS for BK systems to prevent the deployment of ransomware infection. Notably, in case of firewall failure, the Ransomware executables cannot be launched and executed on the BK machine(s) [41].<br>- Using various platforms of virtualization technology (VT) for main and BK servers, e.g. one IntelVT and the other AMD-V, to immune the BK servers against malware able to perform Hyperjacking attacks like Blue Pill. |
| Social Engineering | - The lack of awareness concerning cybersecurity threats among victims is the root cause of the success of social engineering and doxing attacks [77].<br>- A comprehensive understanding of attack vectors is desperately required to develop effective countermeasures and protect employees from social engineering attacks [69].<br>- Cyber-culture and security awareness was recognized as the most effective solution in addressing threats originating from humans. | - Holding regular cybersecurity training programs for personnel of banks, financial institutes, and companies at any level of employment, including high-, mid-, and low-level managers and employees. |
| Shortage of cybersecurity experts | - Training well-skilled cybersecurity experts in the quantity needed in the financial sector requires plenty of experience and practical efforts and cannot be accomplished in a short time [123].<br>- Companies that have hired experts from other areas of computer science as cybersecurity experts have experienced more challenges and suffered more cyber damages as the result of security flaws caused by misconfiguration of cyber defense equipment, in particular IDS, AVs, firewalls, honeypots, and unified threat management (UTM) systems [124], [125]. | - To occupy the cybersecurity specialist posts, employing experts with backgrounds in other areas of computer science and IT rather than pure cybersecurity is strongly negated due to the possibility of misconfiguration of security equipment. |

---

[15] https://www.sophos.com/en-us/press/press-releases/2022/04/ransomware-hit-66-percent-of-organizations-surveyed-for-sophos-annual-state-of-ransomware-2022

| | | |
|---|---|---|
| Lack of CTI Sharing | - Enormous financial loss and reputation damage have been imposed on FinTech-enabled services as the result of electricity disruption and loss of power supplies [61].<br><br>- The majority of financial service providers have restricted access to malicious and abroad IPs. However, this defensive strategy was lacking for many power supplier entities [126]. | - To prevent the propagation of consequences from cyber-attacks against power suppliers to the financial sector, the policy of abroad IP restriction should also be followed by the companies supplying power to the financial industry.<br>- A CTI sharing agreement between the financial and energy sectors is needed to regularly update and sync their threat indicators through common security vendors.<br>- Current legislations, GDPR in the EU and the 1995 Directive in the US, should be revised to fully legalize automated CTI sharing [127]. |
| Lack of cyber-insurance | - Hazards originating from nature, like fire, or human activities like theft, have always caught the attention of investors to arrange insurance policies covering the risks and damages. However, investors and insurance makers have not paid enough attention to hazards originating from cyber incidents. Given many recent cybersecurity incidents resulting in considerable financial loss, cyber insurance policies need to be reconsidered [132]. | - Players in the financial sector are to ensure they have insurance policies to cover cyber incidents, focusing on sufficient responsibilities for cybersecurity attacks, particularly insider attacks. |
| Sluggish Speed of Defenses vs. Threats | - Whereas banks and financial institutions have grown increasingly aware of the changing threat landscape and the debilitating effects of attacks on the availability, integrity, and confidentiality of financial data and services, numerous recent catastrophic cyber-attacks proved that the speed of changing cyber threat landscaper has been much more than the growth in cyber-culture awareness of financial institutes and banks [30]. | - Enhancing the speed of cyber-culture awareness, cybersecurity training, and employing new technology-related solutions like BC.<br>- An immediate update of current defense systems with the latest security patches, specifically for IDS, AVs, and Firewalls. |
| General | - FinTech is a mission-critical application [138]. | - To effectively deal with such a wide range of cyber-attacks threatening the financial sector, the combination of both AI-assisted decision support systems and manual approaches by security experts is indispensable to respond to the sheer number of today's attacks as well as to deal with low and slow APTs and insiders, respectively.<br>- A defense-in-depth (DiD) architecture containing multi-layer defense strategies should be implemented for FinTech services to take effective and on-time actions against imminent threats. |
| General | - It is undeniable in cybersecurity that vulnerabilities and flaws always exist, and no defense strategies can guarantee security to one hundred percent. Hence, it is far more reasonable that these latent vulnerabilities be detected by trusted persons instead of malicious individuals [139]. | - Employing a certified ethical hacking (CEH) team of trusted experts and putting the pre- and post-employment policy.<br>- Performing ethical penetration tests as rigorously as possible against all the assets, equipment, procedures, and even individuals for the big players in the financial sector based on standards and legislation.<br>- OWASP ASVS 4.0.3[16] is one of the most notable penetration test standards for the security verification of web-based applications. |

---

[16] https://owasp.org/www-project-application-security-verification-standard/

## 6.6 Future Research Directions (RQ6)

With the advancement in IT and AI and considering their pervasive nature in all aspects of today's human life, researching to meticulously investigate cybersecurity threats in other sectors, particularly manufacturing, energy, and health, should urgently be undertaken. It paves the way to propose and improve defense solutions against skyrocketed cyber threats that are rapidly evolving. Addressing RQ6, we suggest directions for future research, aiming to address existing drawbacks and limitations with current defensive methods and systems.

### *6.6.1 In terms of Technology*

(a) It was alerted in this paper that, whereas the number of total malware has significantly surged during recent years, the number of detected banking malware has decreased. This proves significant advances in stealth behavior and silent actions in new banking malware generations. Hence, malware detection methods and tools need rapid innovation to detect such intelligent malware classes. To effectively deal with these new technologies that banking malware is equipped with, Anti-malware scanners need to use new hooking technology able to perform deeper scans at the lowest level of the operating systems' kernel. Although deeper hooks make interpreting and modeling the processes' behavior more complex, more stealth malware can be identified and tracked. Using interrupt-level hooks instead of IRP hooks, which are the most common way these days, is our suggestion to be considered by the researchers in the cybersecurity community.

(b) Implementing BC as the infrastructure enables many security advantages. However, the problem of processing big data and today's large-size and high-dimensional datasets in the financial sector caused some obstacles. If these obstacles are tackled, the way will be paved for the rapid adoption of BC in the financial sector. As a short-term research direction, researchers from distributed systems and big data analytics fields could focus on employing BC-powered applications to process big data more effectively with an acceptable response time. Making BC-powered applications adaptable to big real-time systems can also be considered for long-term research.

(c) It is somehow too early to talk about vanishing the concept of money in new supply chains; however, it has been anticipated by some papers studied in this survey that, sooner or later, the concept of money - in its current form - will disappear or be replaced with new euphemisms in future BC-powered supply chains. Studying the feasibility of this paradigm shift can be attractive to researchers and scientists in the fields of the digital economy, information systems, law, and IT as an interdisciplinary research topic.

(d) Although several AC models have been introduced in the literature to guarantee the integrity and confidentiality of information systems in general, none of them concentrated on ensuring the integrity and confidentiality of financial data. The generic AC models impose a considerable delay in granting access to financial data if they want to fully guarantee confidentiality and integrity. Respecting the great importance of the speed of committing financial transactions, the design and implementation of a customized AC system for FinTech rather than generic AC systems can be considered by scholars as an important future research direction, aiming to improve the performance of FinTech systems while integrity and confidentiality are guaranteed.

### *6.6.2 In Terms of Human Factors*

(a) Employing CEH teams to perform penetration testing is a missing piece in the defense solutions offered by papers studied in this survey. Considering today's inter-connected large-scale networks and massive storage of digital information, penetration tests by a trusted CEH team make it feasible for early detection of security flaws and patching them before rivals or malicious hackers can find and exploit them for infiltration, espionage, or destruction. Meanwhile, employing a CEH team needs many sensitive considerations and legal policies besides technical standards, checklists, frameworks, and tools. The investigation and updating of policies and legal affairs required for employing the CEH team is suggested for future research direction, as this evidently lacks sufficient published materials to be referred to.

(b) Because of the misuse of lawful access, detecting insider threats is much more complicated than detecting outsiders, and their repercussions and consequences are far more devastating. Given the hidden nature of insider attacks and their growing trend, future research should consider the rapid development of an insider digital forensic investigation and an insider threat program, as systematic forensic-readiness was discovered to be a neglected point by many entities in the financial sector.

## *6.6.3 In Terms of Procedures*

(a) Whereas some ISO standards were introduced to provide checklists of cybersecurity essentials to be implemented in businesses and enterprises, there is a lack of an ISO standard capable of covering all technology-, human-, and process-based risk factors for the FinTech systems. Hence, we propose that developing such standards and policies, including security checklists for the financial sector, be considered a subject for future research by scientists in the cybersecurity and economy sectors.

(b) Currently, insurance companies offer some insurance policies to cover cyber incidents. However, many ambiguities and controversies exist with these insurance policies, e.g. what kind of cyber incidents should be covered? How to designate the origin of a cyber incident? How to prove whether a cyber incident results from a cyber-attack or not? What kinds of evidence are required to prove the impact of a certain attack? Responding to these concerns requires determining if a cyber insurance policy offers sufficient responsibilities and coverage against cybersecurity attacks. Therefore, we strongly recommend deep and comprehensive reviews of the legal acts to make cyber insurance policies adaptable to cover cybersecurity attacks and digital forensics.

(c) Providing reliable models to measure and quantify cybersecurity risks still has many open challenges. Therefore, much more effort and research need to be conducted to extract a holistic legal framework, as well as in-detail legislation to be cited as a base for measuring cybersecurity risks. Besides, the financial sector needs a dedicated risk management policy as the assets are invaluable, untouchable, and radically different from other domains.

## 7. Conclusion

FinTech has revolutionized the financial sector by delivering global digital services to organizations and individuals. The rapid development of FinTech, further propelled by the COVID-19 pandemic, has brought transformative changes to economies and broadened access to financial services. In this survey paper, we explored the most potential and destructive cybersecurity threats targeting FinTech and reviewed the latest defense solutions, analyzing 74 papers using PRISMA methodology. We introduced a novel taxonomy of security threats in FinTech and elaborated on the application of defensive measures against incoming threats. Our comparison covered various perspectives, including technical details, the impacts, and defense efficacy, shedding light on current shortcomings in threat detection and defense strategy. We proposed actionable insights and recommendations to address these gaps, intending to improve business resilience and sustainability. The findings are of paramount importance to key stakeholders in the financial sector, such as banks and enterprises, as they provide a roadmap for cybersecurity experts to address the ever-evolving threats in this mission-critical arena.

## Data availability

No data was used for the research described in the article.

# Appendix A

**Table A-1:** The list and details of papers included for qualitative synthesis (Step 4) in this survey.

| Threat/Defense | Main category | Subcategory | Reference | Channel | Source | Year |
|---|---|---|---|---|---|---|
| Threats (43 papers) | Technology-based | Malware | [21] | Journal | ScienceDirect | 2018 |
| | | | [24] | Journal | ScienceDirect | 2019 |
| | | | [26] | Journal | ScienceDirect | 2017 |
| | | DDoS | [28] | Journal | IEEE Xplore | 2020 |
| | | | [35] | Journal | ScienceDirect | 2023 |
| | | | [32] | Journal | ScienceDirect | 2018 |
| | | | [33] | Journal | IEEE | 2017 |
| | | Digital Extortion | [37] | Journal | ScienceDirect | 2020 |
| | | | [39] | Journal | IEEE Xplore | 2020 |
| | | | [40] | Journal | IEEE Xplore | 2022 |
| | | | [41] | Journal | IEEE Xplore | 2022 |
| | | Industrial Espionage | [42] | Journal | ScienceDirect | 2020 |
| | | | [43] | Journal | Springer Link | 2020 |
| | | | [44] | Conference | ScienceDirect | 2022 |
| | | Vulnerability in IIOTs | [46] | Journal | ScienceDirect | 2022 |
| | | | [47] | Journal | ACM | 2019 |
| | | | [49] | Journal | ScienceDirect | 2020 |
| | | | [50] | Conference | IEEE Xplore | 2018 |
| | | | [51] | Journal | ACM | 2020 |
| | | | [52] | Journal | ScienceDirect | 2021 |
| | | | [53] | Journal | ScienceDirect | 2022 |
| | | | [54] | Journal | IEEE Xplore | 2021 |
| | | | [55] | Journal | ScienceDirect | 2021 |
| | | Power Attacks | [57] | Journal | ScienceDirect | 2019 |
| | | | [58] | Journal | ScienceDirect | 2020 |
| | | | [59] | Journal | IEEE | 2017 |
| | | | [60] | Journal | ScienceDirect | 2017 |
| | | | [61] | Journal | Springer Link | 2021 |
| | | | [62] | Journal | IEEE Xplore | 2020 |
| | Human-originated | Insider Attacks | [64] | Journal | ScienceDirect | 2022 |
| | | | [65] | Journal | ScienceDirect | 2021 |
| | | Social Engineering | [69] | Journal | ScienceDirect | 2015 |
| | | | [70] | Journal | ACM | 2018 |
| | | | [77] | Conference | ScienceDirect | 2016 |
| | | | [80] | Journal | ScienceDirect | 2021 |
| | | | [81] | Journal | MDPI | 2018 |
| | | Internet Fraud | [83] | Journal | ScienceDirect | 2022 |
| | | | [84] | Journal | ScienceDirect | 2019 |
| | | | [85] | Journal | ACM | 2018 |
| | Procedure-related | Data Leak | [86] | Magazine | IEEE Xplore | 2021 |
| | | | [87] | Conference | ACM | 2020 |
| | | | [89] | Journal | ScienceDirect | 2021 |
| | | Data Breach | [90] | Journal | ScienceDirect | 2022 |
| Defenses (31 + 3* papers) | Technology-based | CTI Sharing | [24]* | Journal | ScienceDirect | 2019 |
| | | | [126] | Journal | ScienceDirect | 2021 |
| | | | [127] | Journal | ScienceDirect | 2017 |
| | | Anti-malware and Intrusion Detection | [39]* | Journal | IEEE Xplore | 2020 |
| | | | [41]* | Journal | IEEE Xplore | 2022 |
| | | | [96] | Journal | IEEE Xplore | 2021 |
| | | | [97] | Journal | ScienceDirect | 2017 |
| | | Access Control Systems | [98] | Journal | ScienceDirect | 2016 |
| | | | [99] | Journal | ScienceDirect | 2018 |
| | | | [100] | Conference | Springer Link | 2019 |
| | | BC Technology | [103] | Journal | ScienceDirect | 2023 |
| | | | [105] | Journal | ScienceDirect | 2021 |
| | | | [107] | Conference | ACM | 2018 |
| | | | [108] | Journal | ScienceDirect | 2021 |
| | | BC Infrastructure — Smart Contracts | [109] | Journal | ScienceDirect | 2021 |
| | | | [110] | Journal | ScienceDirect | 2019 |
| | | | [113] | Journal | Taylor & Francis | 2019 |
| | | | [114] | Journal | Taylor & Francis | 2018 |

|  |  |  | [116] | Journal | Emerald | 2019 |
|  |  | Money Transfer | [117] | Journal | ScienceDirect | 2022 |
|  |  |  | [118] | Journal | Taylor & Francis | 2020 |
|  |  |  | [120] | Journal | ScienceDirect | 2022 |
|  |  | Supply Chain Management | [121] | Journal | ScienceDirect | 2022 |
|  |  |  | [122] | Journal | ScienceDirect | 2022 |
|  | Human-originated | Cyber Culture and Security Awareness | [30] | Journal | ScienceDirect | 2021 |
|  |  |  | [81] | Journal | MDPI | 2021 |
|  |  | Cybersecurity Training | [123] | Journal | ACM | 2018 |
|  |  |  | [124] | Journal | ScienceDirect | 2022 |
|  |  |  | [125] | Journal | Springer Link | 2022 |
|  | Procedure-related | Risk Management Policies | [23] | Journal | Hindawi | 2019 |
|  |  |  | [131] | Journal | ScienceDirect | 2022 |
|  |  | Cyber-insurance | [132] | Journal | ScienceDirect | 2023 |
|  |  | RegTech | [133] | Journal | ScienceDirect | 2022 |
|  |  |  | [134] | Journal | Springer Link | 2018 |

\* These references have studied a threat and a defense approach simultaneously. Hence, they have been listed in both threats and defenses.